\documentclass{aa}  

\usepackage{graphicx}
\usepackage{txfonts}
\usepackage[]{lineno}
\bibpunct{(}{)}{;}{a}{}{,}
\usepackage{float}
\usepackage{hyperref}
\hypersetup{
    colorlinks=true,
    linkcolor=blue,
    filecolor=magenta,      
    urlcolor=blue,
    citecolor=blue,
}
\usepackage{color}

\begin{document} 
\nolinenumbers

   \title{Early-time $\gamma$-ray constraints on cosmic-ray acceleration in the core-collapse SN 2023ixf with the \textit{Fermi} Large Area Telescope}
   \titlerunning{Early-time constraints on cosmic-ray acceleration in SN 2023ixf}

   \author{G. Martí-Devesa
          \inst{1,2}
          \and
          C.~C. Cheung \inst{3}
          \and
          N. Di Lalla \inst{4}
         \and
          M. Renaud \inst{5}
          \and
          G. Principe \inst{2,6,7}
          \and
          N. Omodei  \inst{4}
          \and 
          F. Acero \inst{8,9}
          }

   \institute{Institut f\"{u}r Astro- und Teilchenphysik, Universit\"{a}t Innsbruck, A-6020 Innsbruck, Austria\\\email{guillem.marti-devesa@ts.infn.it}
        \and
        Dipartimento di Fisica, Universit\'a di Trieste, I-34127 Trieste, Italy
                \and
                Space Science Division, Naval Research Laboratory, Washington, DC 20375, USA
                \and
                 W.W. Hansen Experimental Physics Laboratory, Kavli Institute for Particle Astrophysics and Cosmology, Department of Physics and SLAC National Accelerator
Laboratory, Stanford University, Stanford, CA 94305, USA
                \and
                Laboratoire Univers et Particules de Montpellier (LUPM), Universit\'{e} de Montpellier, CNRS/IN2P3, CC72, Place Eug\`{e}ne Bataillon, F-34095 Montpellier
Cedex 5, France
        \and
                Istituto Nazionale di Fisica Nucleare, Sezione di Trieste, 34127 Trieste, Italy
        \and
        INAF - Istituto di Radioastronomia, I-40129 Bologna, Italy
        \and
        Universit\'e Paris-Saclay, Universit\'e Paris Cit\'e, CEA, CNRS, AIM, 91191, Gif-sur-Yvette, France
        \and
        FSLAC IRL 2009, CNRS/IAC, La Laguna, Tenerife, Spain\\
}

   \date{Received ---, ---; accepted ---, ---}

% \abstract{}{}{}{}{} 
% 5 {} token are mandatory
  \nolinenumbers
  \abstract
   {While supernova remnants (SNRs) have been considered the most relevant Galactic cosmic ray (CR) accelerators for decades, core-collapse supernovae (CCSNe) could accelerate particles during the earliest stages of their evolution and hence contribute to the CR energy budget in the Galaxy. Some SNRs have indeed been associated with TeV $\gamma$-rays, yet proton acceleration efficiency during the early stages of an SN expansion remains mostly unconstrained.}
   {The multi-wavelength observation of SN 2023ixf, a Type II supernova (SN) in the nearby galaxy M101 (at a distance of 6.85 Mpc), opens the possibility to constrain CR acceleration  within a few days after the collapse of the red super-giant stellar progenitor. With this work, we intend to provide a phenomenological, quasi-model-independent constraint on the CR acceleration efficiency during this event at photon energies above 100 MeV.}
   {We performed a maximum-likelihood analysis of $\gamma$-ray data from the \textit{Fermi} Large Area Telescope up to one month after the SN explosion. We searched for high-energy, non-thermal emission from its expanding shock, and estimated the underlying hadronic CR energy reservoir assuming a power-law proton distribution consistent with standard diffusive shock acceleration.}
   {We do not find significant $\gamma$-ray emission from SN 2023ixf. Nonetheless, our non-detection provides the first limit on the energy transferred to the population of hadronic CRs during the very early expansion of a CCSN. 
   }
   {Under reasonable assumptions, our limits would imply a maximum efficiency on the CR acceleration of as low as 1\%, which is inconsistent with the common estimate of 10\% in generic SNe. However, this result is highly dependent on the assumed geometry of the circumstellar medium, and could be relaxed back to 10\% by challenging spherical symmetry.
   Consequently, a more sophisticated, inhomogeneous characterisation of the shock and the progenitor's environment is required before establishing whether or not Type II SNe are indeed efficient CR accelerators at early times.}
   

   \maketitle
%
%-------------------------------------------------------------------

\nolinenumbers
\section{Introduction} \label{sec:intro}

The origin of cosmic rays (CRs) is still an open issue, and the Galactic sources that can accelerate CRs up to the so-called {knee} of the CR spectrum at PeV energies have yet to be determined. In the standard paradigm (which includes both CR production and transport), the bulk of Galactic CRs would be produced in supernovae (SNe) and their remnants \citep[SNRs; see][for a review]{Blasi13}. In such a scenario, it is particularly relevant that SN events typically release an energy of $\sim 10^{51}$ erg, meaning that, at their observed rate, a $\sim 10\%$ energy transfer into CRs is sufficient to explain the energetics of Galactic CRs.

However, and among other issues \citep[see e.g.][]{Gabici19}, very-high energy (VHE; $>100$ GeV) observations with Cherenkov telescopes consistently reveal multi-TeV cut-offs in most SNRs \citep[see e.g.][]{Acero15, Ahnen17, HESS18}. That is, known Galactic SNRs do not appear to be significant contributors to the CR flux at PeV energies at their current evolutionary stage. This inconsistency with the standard paradigm can be alleviated if SNe accelerate CRs up to these energies at very early stages \citep[i.e. within a few days after the event; ][]{Voelk88, Tatischeff09, Bell13, Schure13, Bykov18, Marcowith18, Cristofari20, Inoue21, Brose22}. However, testing this hypothesis remains challenging due to the low number of SNe detected at sufficiently short distances. Some studies have attempted to obtain such constraints before \citep{Margutti14, Ackermann15, HESS15, MAGIC17, HESS19, Murase19, Prokhorov21}, but obtaining effectual limits on the CR population at early times is inherently problematic. On one hand, TeV facilities can rapidly observe a SN within days after the outburst, but $\gamma$--$\gamma$ absorption heavily impacts the expected flux at those times \citep{Marcowith14}. On the other hand, current GeV $\gamma$-ray detectors require long exposure times ($\sim1$~month) to reach noteworthy luminosity limits for distant events \citep{Ackermann15}. Hence, previous efforts to estimate the CR energy reservoir during the expansion of SNe were impacted by such limitations.

SN 2023ixf is a SN Type II \citep[for a review see e.g.][]{Smith14} recently discovered by \cite{Itagaki23} in the nearby M101 \citep[$D=6.85$ Mpc;][]{Riess22}, with an estimated explosion time at $T_0 = 60082.743\pm0.083$ MJD \citep{Hiramatsu23}. Its progenitor was a red supergiant \citep[RSG; see e.g.][]{Jencson23, Kilpatrick23, Niu23, Qin23, Xiang23}, with an increased mass-loss rate during the last years before the outburst of $\dot{M}_{\rm RSG} = 10^{-3}$--$10^{-2}$ M$_{\odot}$/yr \citep{Bostroem23, Hiramatsu23, Jacobson-Galan23, Teja23}. Together with a large shock and wind velocities inferred \citep[$V_{s, 0}\simeq10^4$ km/s and $u_w\simeq100$ km/s;][]{Smith23}, this leads to the ideal conditions for particle acceleration and $\gamma$-ray production through hadronic channels. Together with its proximity, $\gamma$-ray observations of SN 2023ixf thus offer an unprecedented opportunity to explore CR acceleration in the very early expansion of the SN shock.

Here we report a quasi-model-independent constraint on the CR acceleration in SN 2023ixf, providing an experimental test to the hypothesis that CCSNe  efficiently accelerate protons at early times. In Section \ref{sec:data} we present the GeV $\gamma$-ray observations and data analysis procedure, with the corresponding results detailed in Section \ref{sec:results}. Through Section \ref{sec:discussion} we estimate the corresponding CR acceleration efficiencies, while the possible major biases in our assumptions are discussed in Section \ref{sec:bias}. Finally, we summarise our findings in Section \ref{sec:summary}.

%--------------------------------------------------------------------
\section{Observations and analysis} \label{sec:data}

%-------------------------------------- Two column figure (place early!)

The Large Area Telescope (LAT) is a pair-production $\gamma$-ray detector on board the \textit{Fermi} satellite, launched in 2008 \citep{FermiLAT}. Operating in an all-sky survey mode, it detects $\gamma$-ray photons from 20 MeV to more than 500 GeV. As it observes the whole sky every $\sim 3$ h, \textit{Fermi}-LAT is an ideal instrument to detect and follow-up on high-energy transient sources. 

The LAT point-spread function (PSF) is highly energy-dependent, being able to reconstruct the incoming direction of $\gamma$-rays at $68\%$ confidence level within $5^{\circ}$ and $0.1^{\circ}$ for $100$ MeV and $>10$ GeV events, respectively. We note that at even lower energies, the large PSF might lead to source confusion between point-like sources \citep[for an alternative analysis implementation, see][]{Principe18}. Therefore, for our standard analysis we select only P8R3 data \citep{Atwood13,P8R3} between 100 MeV and 500 GeV within a region of interest (ROI) of $10^{\circ} \times 10^{\circ}$ centred at the optical position of SN 2023ixf \citep[Right Ascension (RA) = 14:03:38.580, Declination (Dec) = +54:18:42.10;][]{Itagaki23}. To study the early expansion of the SN shock, we selected data from $T_0$ up to 31 days after the outburst (mission elapsed time (MET); 706125000 -- 708803400). We applied a maximum zenith angle cut at $90^{\circ}$ to reduce Earth--limb contamination while averaging over the azimuthal angle, and selecting only FRONT+BACK SOURCE events (evtype=3, evclass=128). In addition, a DATA$\_$QUAL>0 and LAT$\_$CONFIG==1 filter is applied. For its subsequent analysis, the ROI is then divided into $0.1^{\circ}$ spatial bins, while using eight bins per logarithmic energy decade.

\begin{figure}
  \centering
   \includegraphics[width=\columnwidth]{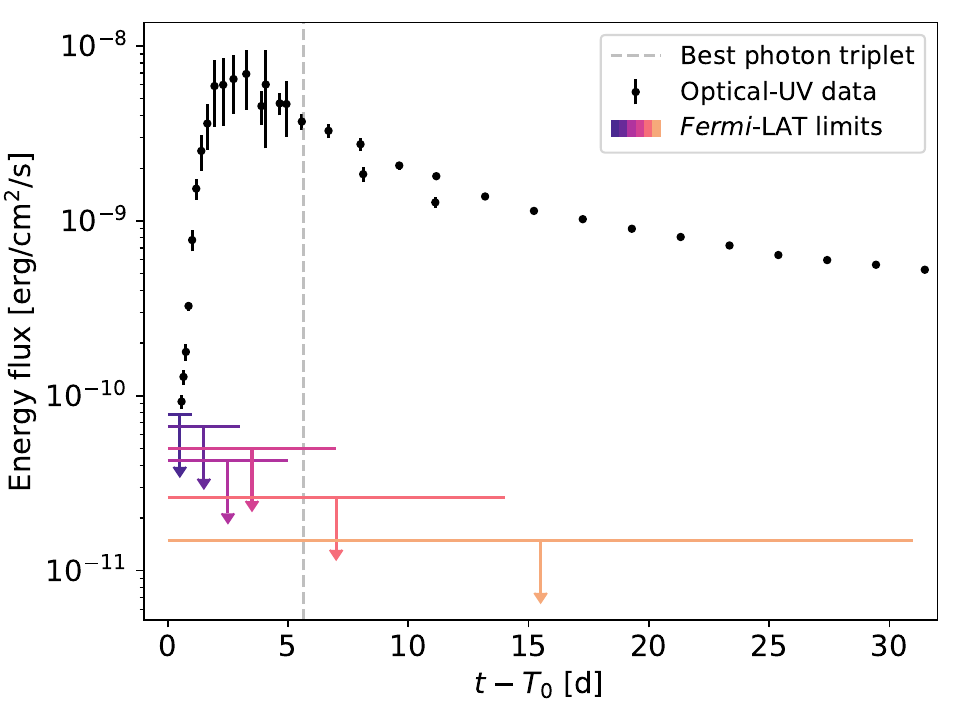}
   \caption{\textit{Fermi}-LAT upper limits on the integrated energy flux above 100~MeV for different exposure times, compared with the best-fit optical energy flux from \citet{Zimmerman23} in black. We note that observations for $\Delta T = 7$ d are less constraining than $\Delta T = 5$ d as a result of having a  larger $TS$ value, although they are still consistent with a statistical fluctuation (see Appendix \ref{A:limits}). The grey vertical line indicates the arrival time of the best photon triplet candidate.}
     
         \label{Fig:eflux}
  \end{figure}

We performed a maximum-likelihood analysis on this dataset \citep{Mattox96} using \texttt{fermipy} and the \texttt{fermitools} \citep{Fermipy, Fermitools}. As our background model, we employed the third 4FGL data release (4FGL-DR3)
---which is based on 12 years of survey data \citep[][]{4FGL, 4FGL-DR3}--- up to $2.5^{\circ}$ beyond the edge of our ROI, as well as the latest Galactic and isotropic diffuse models (\texttt{gll\_iem\_v07.fit} and \texttt{iso\_P8R3\_SOURCE\_V3\_v1.txt})\footnote{\url{https://fermi.gsfc.nasa.gov/ssc/data/access/lat/BackgroundModels.html}}. To evaluate the detection significance of individual sources, we employed the test statistic ($TS$) defined as

\begin{equation}
TS = -2\;\rm{ln} \left( L_{0}/L_{1}\right) \; ,
\end{equation}

\noindent where $L_0$ is the likelihood value for the null hypothesis and  $L_1$ the likelihood for the complete model. The $TS$ follows a $\chi^2$ distribution, and the larger its value, the less likely the null hypothesis. As an example, for one degree of freedom, $\sqrt{TS}$ will approximately be the resulting significance in sigma ($\sigma$).

\begin{figure*}
  \centering
   \includegraphics[width=\columnwidth]{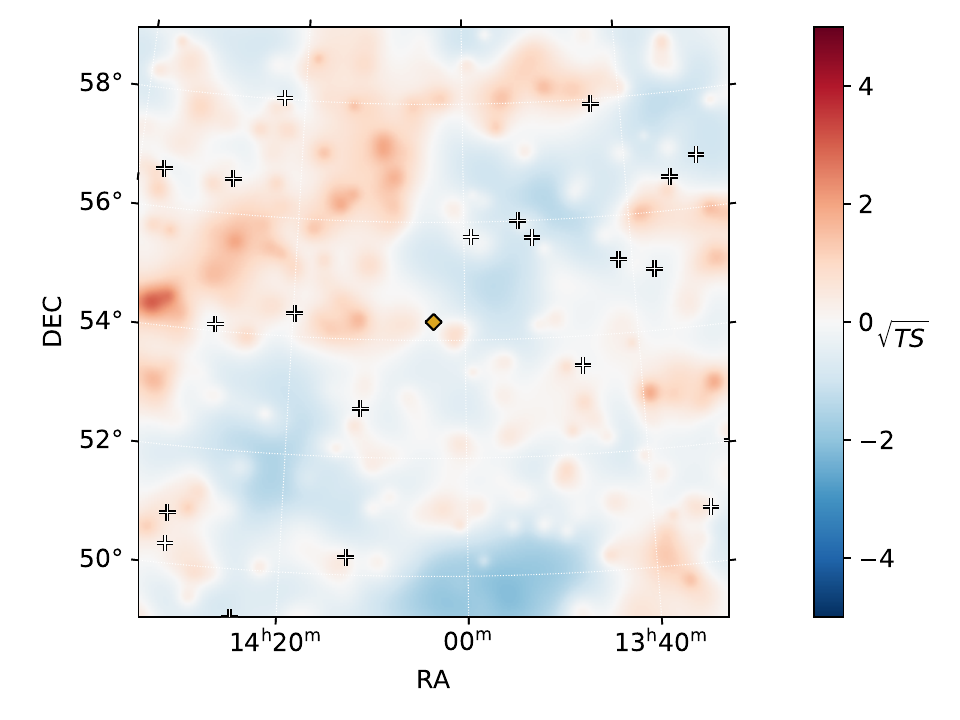}
   \includegraphics[width=\columnwidth]{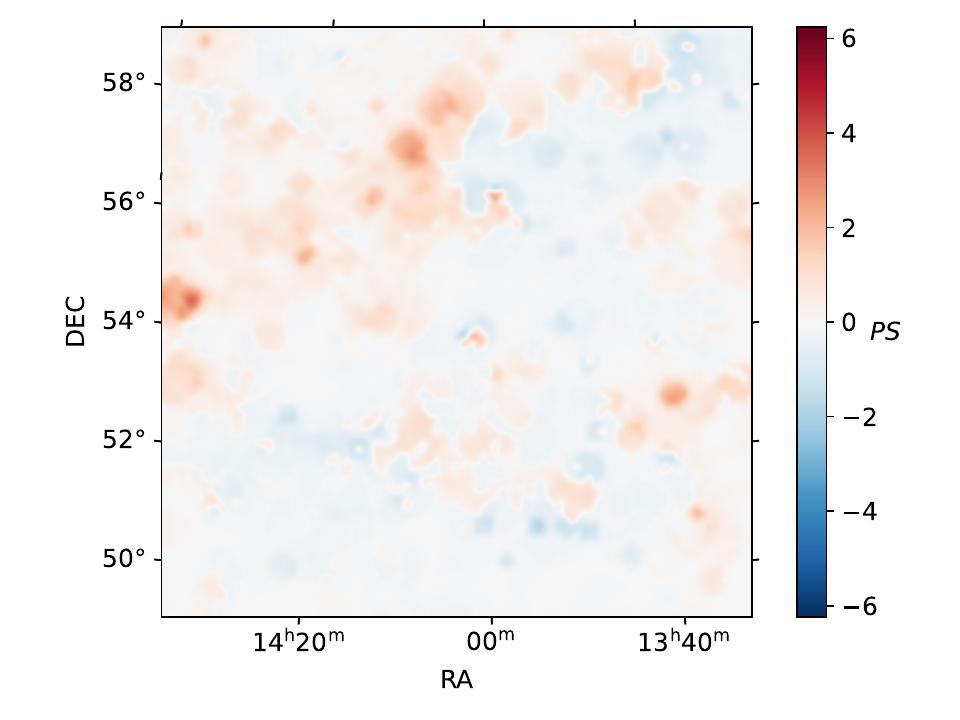}
   \caption{\textit{Fermi}-LAT smoothed residual map with a bicubic interpolation of our ROI centred on SN 2023ixf for 1 month of observations after $T_0$. (\textit{Left}): Standard significance map as implemented in \textit{fermipy} \citep{Fermipy}. White crosses represent 4FGL sources included in the background model, while the yellow diamond is the nominal position of SN 2023ixf. (\textit{Right}): Data--model deviation estimate employing a \textit{PS} map of the same ROI saturated at $PS=6.24$ \citep[equivalent to a $5\sigma$ threshold;][]{Bruel21}.}
     
         \label{Fig:residual}
  \end{figure*}

First we adjust our background model to our ROI by means of the \texttt{optimize} function in \texttt{fermipy}, and refine it within its central region, fitting the normalisation parameters of all sources within $3^{\circ}$ of SN 2023ixf (including the diffuse components). In a final step, we include an additional point-like source to account for any putative emission from the SN, for which we assume a simple power-law spectral model:

\begin{equation}
                \dfrac{dN_{\rm{\gamma}}}{dE} = N_0\left(\dfrac{E}{E_0}\right) ^{-\Gamma}
,\end{equation}

\noindent with spectral photon index $\Gamma = 2$. In addition to our one-month analysis, the same analysis procedure is applied to exposure times of 1, 3, 5, 7, and 14 days\footnote{These analyses complete our preliminary report in \citet{Marti-Devesa23}.}. In each data set, we correct for energy dispersion for all sources, except for the isotropic diffuse emission, which does not require it. 

\section{Results } \label{sec:results}

Our analysis from Section \ref{sec:data} does not significantly detect SN~2023ixf; therefore, here we report the $\gamma$-ray limits obtained.

\subsection{Maximum-likelihood flux limits}

As no significant flux is obtained in any of the time windows explored, we provide 95\% confidence upper limits on the integrated photon and energy fluxes above 100 MeV (Fig. \ref{Fig:eflux}), and derive spectral energy distributions with two logarithmic energy bins per decade. A complete summary of the limits for different times and energies is provided in Appendix \ref{A:limits}.

For a distance $D=6.85$ Mpc, this corresponds to a luminosity limit of $L_{\gamma}(>100\rm{\;MeV})< 8.4\times10^{40}$ -- $4.8\times10^{41}$ erg/s, depending on the exposure time. This is further complemented with light curves derived with bins of 1, 3, and 7 days in size. In each bin, the normalisations of the five brightest sources in the ROI are first left free, and then the normalisation of SN 2023ixf is estimated. These analyses also do not report significant emission in any time bin, nor the presence of any flare at later times. The largest significance is obtained on the 3rd and 24th days after $T_0$, with $TS=4.8$ and $TS=4.9$, respectively ($\sim 2\sigma$, consistent with a statistical fluctuation). Similarly, no significant excess is found at other positions of the ROI. For completeness, the residual map obtained for the one-month analysis is shown in Fig. \ref{Fig:residual}, and cross-checked with the p-value statistic ($PS$) data--model deviation estimator developed by \citet{Bruel21}. A $\sim 3\sigma$ excess is appreciable at the left edge of the ROI with both methods, far away from the SN.

Alternatively, we can also explore short timescales around the SN explosion. The position of SN 2023ixf was not visible by the LAT at $t = T_0$ but entered the LAT field of view at $T_0 + 4.8$~ks and remained observable until $\sim T_0 + 8$~ks. In analogy to what is commonly done for the analysis of $\gamma$-ray bursts with the LAT, we also performed an unbinned maximum-likelihood analysis with \textit{gtburst}\footnote{\url{https://fermi.gsfc.nasa.gov/ssc/data/analysis/scitools/gtburst.html}} over this time interval using P8R3 data with TRANSIENT10E event class. We selected photons with energies of between 100~MeV and 500~GeV within an ROI with a radius of 12$^{\circ}$ centred on the optical position of SN 2023ixf and with a maximum zenith angle of 100$^{\circ}$. As in the previous analysis, we included the background contribution from sources of the 4FGL catalogue as well as the latest Galactic and isotropic diffuse models (\texttt{gll\_iem\_v07.fit} and \texttt{iso\_P8R3\_TRANSIENT010E\_V3\_v1.txt}). Once again, we find no significant detection at the SN location ($TS=2.0$) and, assuming a power-law model with a photon index of $\Gamma = 2$, we set a 95\% confidence upper limit on the integrated energy flux above 100 MeV of $5.2 \times 10^{-10}$~erg cm$^{-2}$ s$^{-1}$.

\subsection{Photon clusters search} \label{sec:clusters}
Given the lack of significant $\gamma$-ray emission with the standard likelihood analysis (see Section \ref{sec:data}), we searched for individual photons possibly connected to the SN event. To this aim, we applied an analysis method to search for photon triplets
\cite[for a detailed description see][]{2021NatAs...5..385F}, which has previously been adopted in searches for $\gamma$-ray signals from magnetar flares and fast radio bursts \citep{2021NatAs...5..385F, 2023A&A...675A..99P}.

\begin{figure}
  \centering
   \includegraphics[width=\columnwidth]{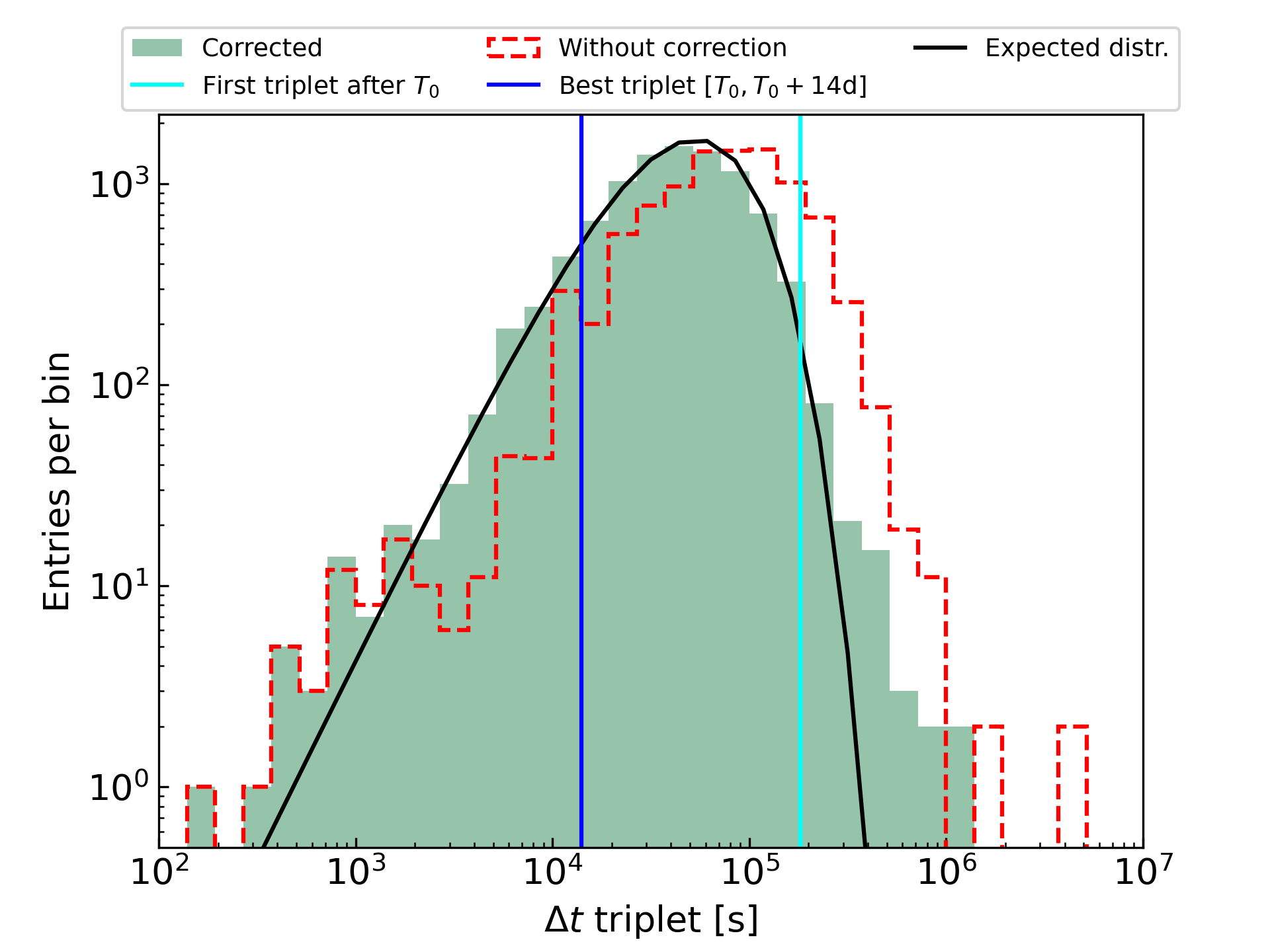}
   \caption{Triplet distribution. This is the distribution of the time intervals $\Delta t$ of the photon triplets with (filled green) and without (red line) considering the correction of the bad intervals due to the LAT orbit and field of view. The expected distribution for independent events is shown as a black line. The vertical lines represent the first photon triplet after $T_0$ (in cyan) and the shortest duration triplet in the two weeks after $T_0$ (in blue), respectively. The latter is obtained from an analysis of a two-week interval after $T_0$ in order to account for the uncertainty on the estimated $T_0$, as well as for a possible delay in the $\gamma$-ray emission from the SN. A significant triplet would appear in the left tail of the distribution. This is not the case for either of the highlighted triplets.}
         \label{Fig:triplets}
  \end{figure}

For our analysis, we selected all the SOURCE class photons between 100 MeV and 500 GeV detected by the LAT over 14.9 years (i.e. from the start of operations until June 30, 2023) in an ROI with a radius of 1$^{\circ}$ centred on the source position. We estimated the time interval $\Delta t_i$ for each triplet of photons $i$ formed by three consecutive events,
\begin{equation}
    \Delta t_i = t_{i+2} - t_i,
\end{equation}
\noindent and corrected this quantity for the effect of bad time intervals by subtracting, from each $\Delta t_i$, the period of time during which the ROI was not observable by the LAT.

In addition to creating the distribution of the whole photon triplets coming from the source position, we investigated the first triplet after $T_0$ and its potential association to the SN event, as well as the shortest duration triplet in the two weeks after $T_0$ for possible long-term emission.

Similarly to the procedure used in \citet{2021NatAs...5..385F} and \citet{2023A&A...675A..99P}, we used the likelihood-ratio method defined in \citet{1983ApJ...272..317L} to estimate the probability that a cluster of three photons occurs by chance due to statistical fluctuations of the background, in the time range $\Delta t_\mathrm{SN} = t_\mathrm{\gamma-ray,3} - T_\mathrm{T0}$, where $t_\mathrm{\gamma-ray,3}$ is the time of the third photon of the triplet after $T_0$. For further details on the derivation of significance of the photon triplets following $T_0,$ see Eqs. 4 and 5 in \citet{2023A&A...675A..99P}.

Figure \ref{Fig:triplets} shows the distribution of the photon triplets. The first photon triplet after $T_0$ is detected on 2023 May 18, at 20:45 UTC (about 3 hours after $T_0$) and presents a duration of $\Delta t \sim 181900$ s.
This triplet presents a pretrial p-value of smaller than 2$\sigma$, indicating that it is most likely due to a statistical fluctuation.
Considering the uncertainty on the estimated $T_0$, as well as a possible delay on the $\gamma$-ray emission from the SN, we also searched for the shortest duration photon triplet in the 14 days after $T_0$. 
The shortest duration triplet ($\Delta t = 14062$ s) was observed at 9:26 UTC on 2023 May 24, more than five days after the SN event and past the optical peak (Fig. \ref{Fig:eflux}). In this case, even if we would expect a flash of $\gamma$-ray emission from the SN at the time of the first photon of the triplet, the probability of these three photons to be associated to the SN is smaller than $2.4\sigma$.

\section{Discussion } \label{sec:discussion}

\begin{figure}
  \centering
   \includegraphics[width=\columnwidth]{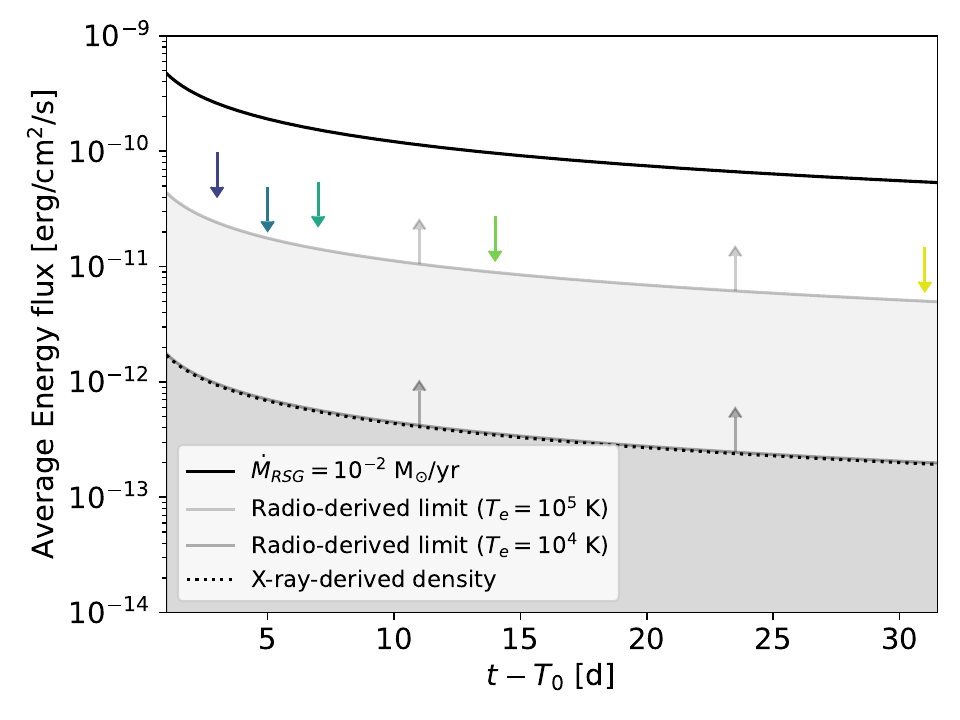}
   \caption{Average integrated energy flux for SN 2023ixf as predicted by the SN 1993J model from \cite{Tatischeff09} compared with integrated energy flux upper limits from the \textit{Fermi}-LAT (coloured markers). Fluxes and limits are integrated starting at $T_0+1$ d. We assume $u_w = 100$ km/s, in line with spectroscopic results \citep{Smith23}. Radio/millimetre (230 GHz) lower limits from \cite{Berger23} ($u_w = 115$ km/s) for free-free absorption assuming different electron temperatures $T_e$ are also displayed. Finally, the prediction derived from X-ray absorption features in \cite{Grefenstette23} (marginally consistent with millimetre-wavelength limits) is shown for completeness. }
     
         \label{Fig:Light_curve}
  \end{figure}

We can firstly assess the relevance of the limits derived above by comparing them with non-thermal expectations from similar SNe. For this purpose, we briefly consider here the model from \citet{Tatischeff09} developed for SN 1993J in M81. This was also a nearby SN \citep[$\sim 3.4$ Mpc;][]{Kudritzki12} with a RSG progenitor. In \cite{Tatischeff09}, the radio emission from the expanding shell is modelled by assuming diffusive shock acceleration (DSA) as the dominant acceleration mechanism, but also including non-linear effects. The differential $\gamma$-ray flux expected for SN 2023ixf at early times would then be \citep{Cristofari20}:

\begin{multline}
\frac{dN_{\gamma}}{dE} = 3.5 \times 10^{-11} \left[\frac{D}{6.85 \rm{\;Mpc}} \right]^{-2} \left[\frac{\dot{M}_{\rm RSG}}{ 10^{-2} \rm{\;M_{\odot}/yr}} \right]^{2}  \left[\frac{t}{1 \rm{\;d}} \right]^{-1} \\ \times \left[\frac{V_{\rm{s}}}{10^4 \rm{\;km/s}} \right]^{2} \left[\frac{u_{\rm{w}}}{100 \rm{\;km/s}} \right]^{-2} \left[\frac{E}{1 \rm{\;TeV}} \right]^{-2} \rm{TeV}^{-1} \rm{cm}^{-2} \rm{s}^{-1}.
\end{multline}

\begin{figure}
  \centering
   \includegraphics[width=\columnwidth]{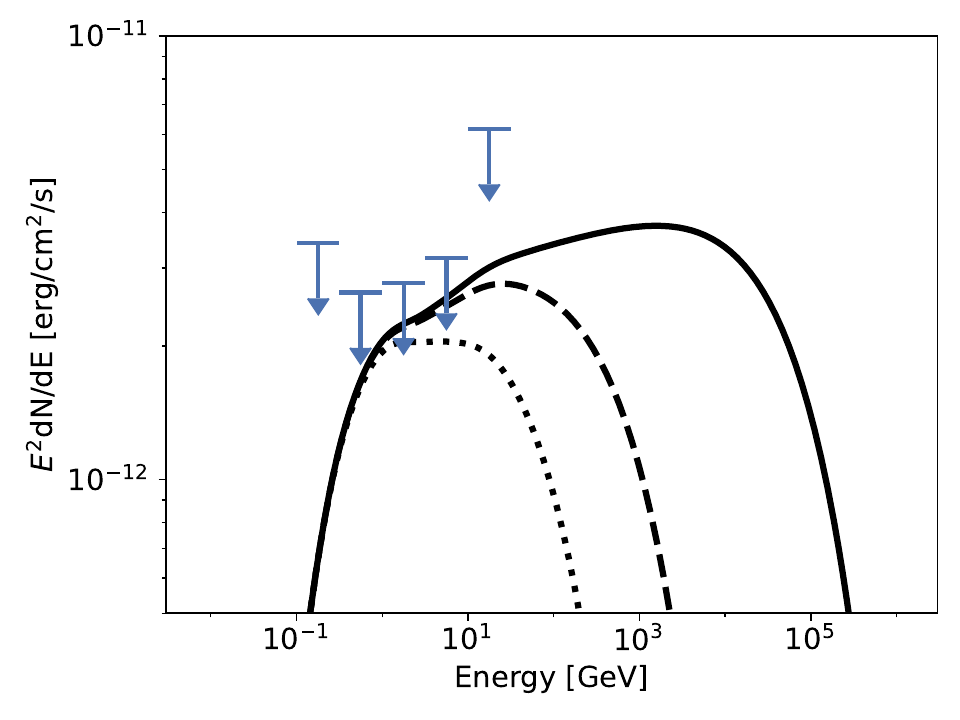}
   \caption{\textit{Fermi}-LAT limits at 95\% confidence level for the one-month time interval compared with the $\pi^0$ decay differential flux from a proton population with $E_{\rm cutoff}=1$ PeV (solid line), 10 TeV (dashed line), or 1 TeV (dotted line). These have the same normalisation $N_0$, and a CR proton energy of $E_{\rm CR} =2.7\times 10^{46}$ erg, $E_{\rm CR} =1.8\times 10^{46}$ erg, and $E_{\rm CR} =1.5\times 10^{46}$ erg, respectively. Limits at energies greater than $10^{1.5}$ GeV lie above $10^{-11}$ erg/cm$^2$/s. For the target material, a density $\rho_0 =5.6\times 10^{-14}$ g/cm$^3$ \citep{Bostroem23} is assumed.}
     
         \label{Fig:SED_example}
  \end{figure}

\noindent Using the inferred values from multi-wavelength observations of SN 2023ixf (Appendix \ref{B:ratios}), we can compute the expected average integrated energy flux between 100 MeV and 100 GeV for different exposure times after $t_i = T_0$+1 d. The shock velocity assumed ($V_s = 10^4$ km/s) is also typical for a Type II SN, and is consistent with the limits from observational constraints on SN 2023ixf \citep{Bostroem23, Jacobson-Galan23, Teja23}. As we cannot directly compare the prediction with our observational limits as derived in Section~\ref{sec:data}, we repeat the analyses including data up to the same dates ($t_f = T_0+$ 3, 5, 7, 14, and 31 days), but starting at $t_i$. In Fig. \ref{Fig:Light_curve} we observe that the prediction lies well above our upper limits for a representative mass-loss rate inferred from optical observations. We note that the LAT limits are derived for a flat ($\Gamma=2$) spectrum, while realistically a hadronic spectrum will be mildly different, most notably with a break at the lowest energies (the so-called $\pi^0$-bump, which we consider in Section \ref{sec:ECR_estimate}). Neglecting a realistic hadronic scenario in Fig. \ref{Fig:Light_curve} leads to overestimation of the upper limits on the integrated energy flux by $\sim10\%$, thus not affecting our conclusion.

Combined with the results from the Submillimeter Array (SMA) at 230 GHz \citep[][]{Berger23}, our limits exclude a substantial fraction of the parameter space for the direct application of the underlying assumptions in the SN 1993J model to SN 2023ixf. Therefore, LAT observations are able to directly constrain the efficiency of the system at transferring kinetic energy of the expanding shock into high-energy CRs. For simplicity, we did not construct a complete model for the shock in SN~2023ixf, but instead estimated the CR energy reservoir directly assuming that protons can gain energy through DSA. This process leads to an average differential proton distribution that at a certain time $t$ follows a power law with an exponential cut-off in momentum space; that is,

\begin{equation}
                \dfrac{dN_{\rm{p}}}{dE} = \beta N_0\left(\dfrac{E}{E_0}\right) ^{-p} \exp{-\left(\frac{E}{E_{\rm cutoff}}\right)} \;,
  \label{eq:PLexpcut}
\end{equation}

\noindent where $\beta$ is the particle's velocity, which corrects the spectral shape of a proton population below $\sim 1$~GeV when described in kinetic energy \citep{Dermer12}. Standard DSA predicts that $p = 2$ \citep[see e.g.][]{Blasi13}, and as our goal is to see if SNe can accelerate protons (at least) up to the knee of the CR spectrum, we fix $E_{\rm cutoff}$ to 1 PeV.
 Tentatively, this should be the most efficient acceleration mechanism at the shock. As most non-linear considerations would effectively soften the spectrum, we note that in order to maintain the same proton density at 1 PeV, the inclusion of those effects does increase the fluxes expected in the GeV band, making any constraints even more stringent. Hereafter, we also set a minimum proton kinetic energy of 100 MeV.

\subsection{Limits on the total energy transferred into CRs} \label{sec:ECR_estimate}

The accelerated protons will interact with the gas in their immediate surroundings, leading to hadronic cascades and a subsequent $\gamma$-ray production; for example, through $\pi^0$ decays.

\begin{figure}
  \centering
   \includegraphics[width=\columnwidth]{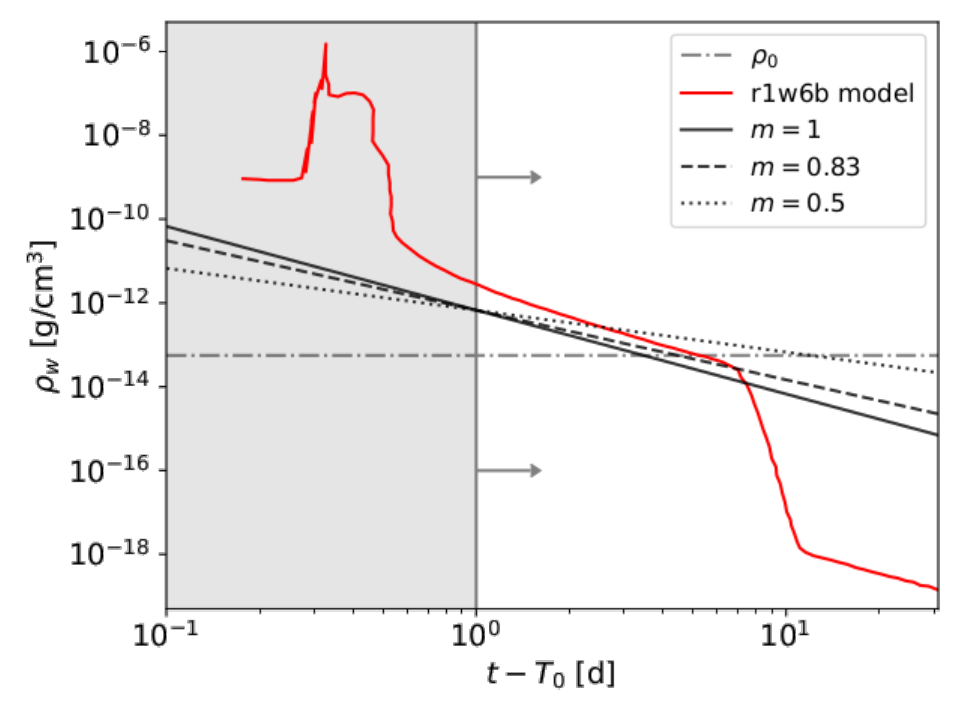}
   \caption{Density profile encountered by the shock as a function of time for different expansion parameters $m$ \citep[including $m=0.83$ as in][a model applied to SN 1993J]{Tatischeff09}. We assumed $\dot{M}_{\rm RSG} = 10 ^{-2}$ M$_{\odot}$/yr, $u_w = 100$ km/s, and $V_{s, 0}=10^4$ km/s. The average density $\rho_0$ obtained by \cite{Bostroem23} is also displayed, as well as the best-fit density profile from the r1w6b model in \citet{Jacobson-Galan23}. The grey-shaded region excludes the times not considered in our discussion on the SN energy budget.}
     
         \label{Fig:rho_profile}
  \end{figure}

We can compute the expected $\gamma$-ray spectral energy distribution (SED) for a proton population that follows the distribution in Eq. \ref{eq:PLexpcut} for any arbitrary $N_0$ given a target proton density (see Fig. \ref{Fig:SED_example}). For this purpose, we used the package \texttt{naima} \cite[\cite{Zabalza15}, which incorporates the cross-section $\sigma _{pp}$ from][]{Kafexhiu14} to compute the SED for several proton populations. In particular, we computed $\gamma$-ray fluxes expected from ten CR distributions that have total energies $E_{CR}$ of between $10^{45}$ and $10^{48}$ erg (uniformly distributed in log $E_{\rm CR}$). Furthermore, and given that the $\gamma$-ray flux depends linearly on the target density, we normalised the results to the average density of $\rho_0 =5.6\times 10^{-14}$ g/cm$^3$ as derived in \citet{Bostroem23} for a compact hydrogen-rich, ionised medium surrounding the progenitor (see Fig. \ref{Fig:rho_profile}). We consider $\rho_0$ to be a conservative reference value, as \citet{Zimmerman23} report a local density of $\rho \sim 5 \times 10^{-13}$ g/cm$^3$. Assuming a proton spectrum with $p > 2$ increases the flux at lower energies (0.1--1~GeV) for any arbitrary $E_{CR}$ normalisation, and hence limits derived for $p=2$ should be considered conservative. Those predictions do not consider absorption processes, as they are likely negligible at MeV and GeV energies (see Section~\ref{sec:absorption}).

\begin{figure*}
  \centering
   \includegraphics[width=\columnwidth]{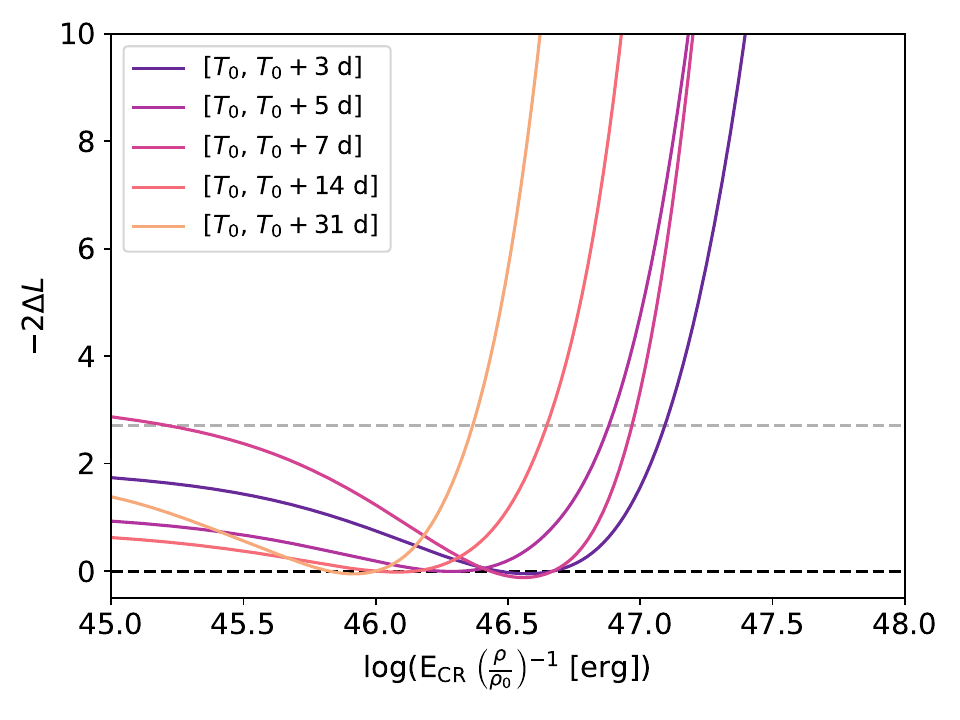}
   \includegraphics[width=\columnwidth]{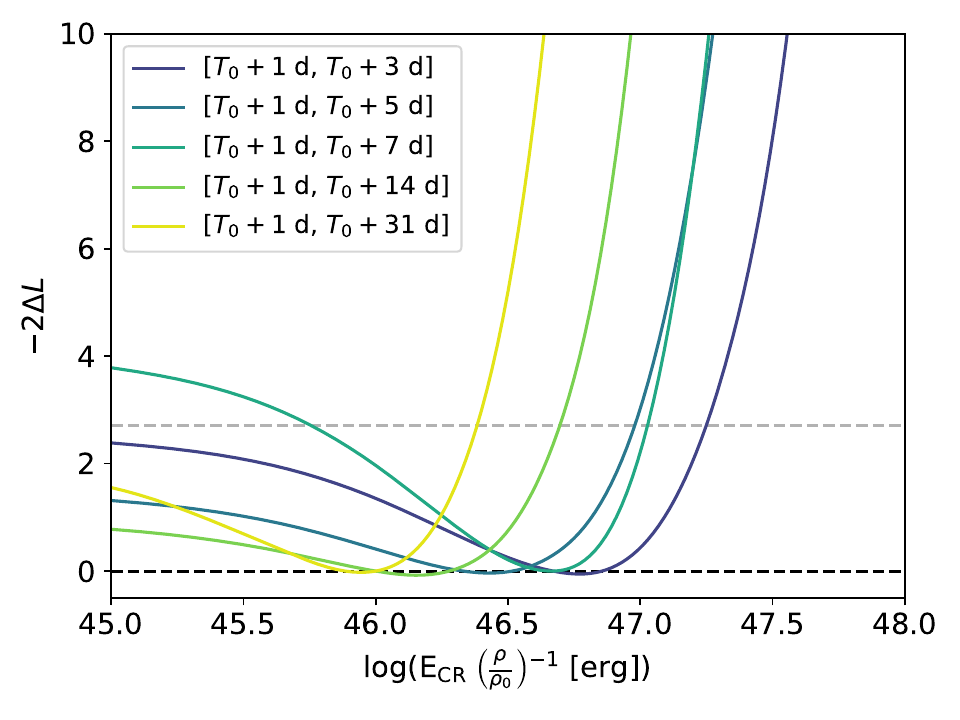}
   \caption{Likelihood profiles assuming a flat $\rho_0$ density profile for different exposure times $\Delta T$, starting at either $T_0$ (\textit{left}) or $T_0+1$ d (\textit{right}). $\Delta L$ is defined as the difference between the likelihood and that corresponding to the $E_{\rm CR}$ value which maximises it unconditionally (black dashed line). In all panels, the grey dashed line represents the 95\% limit, and the obtained likelihoods are interpolated employing a cubic spline.} 
     
         \label{Fig:Likelihood_profile}
  \end{figure*}

Our fitted background models in Section \ref{sec:data} can now be modified to include an additional point-like source at the nominal optical position of SN 2023ixf, but whose SED is that computed through hadronic interactions instead of a simple power law. Therefore, we can then calculate the likelihood for each precomputed CR distribution, and obtain a likelihood profile as a function of $E_{CR}$ normalised at $\rho_0$ for every exposure time (Fig.~\ref{Fig:Likelihood_profile}). The larger $E_{CR}$, the larger the flux, and if it exceeds the sensitivity limit of the LAT, the difference between the newly computed likelihood and its maximum value from our analysis will increase. This allows us to compute 95\% confidence upper limits \citep[corresponding to a $TS$ = 2.706 for a one-sided $\chi^2$ distribution; see e.g.][]{Rolke05} assuming a target density of $\rho_0$.

     \begin{figure*}
  \centering
  \includegraphics[width=\columnwidth]{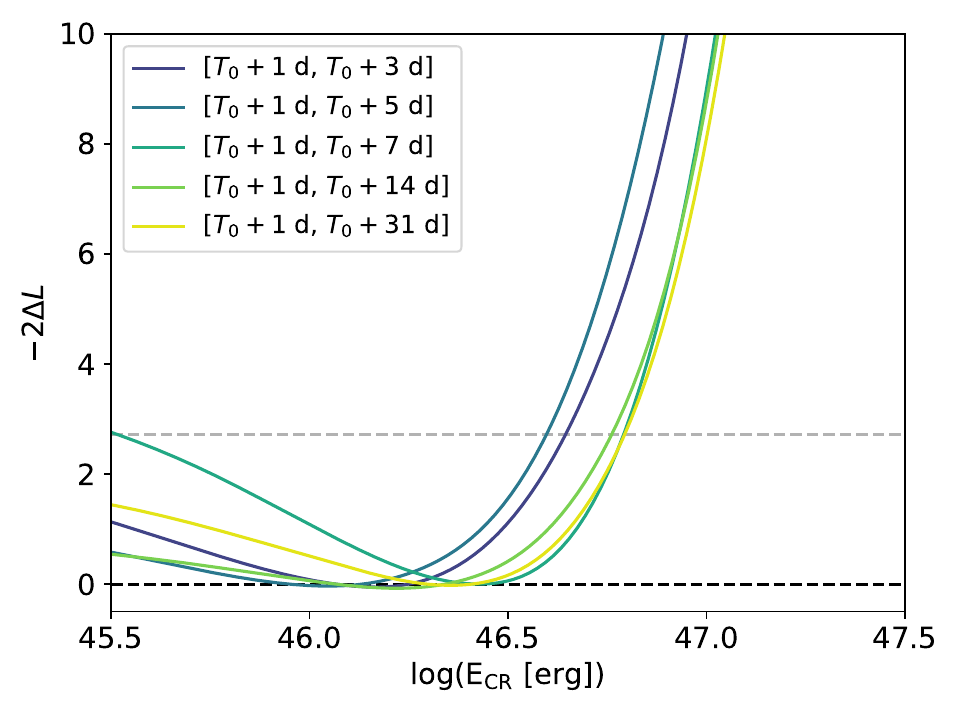}
   \includegraphics[width=\columnwidth]{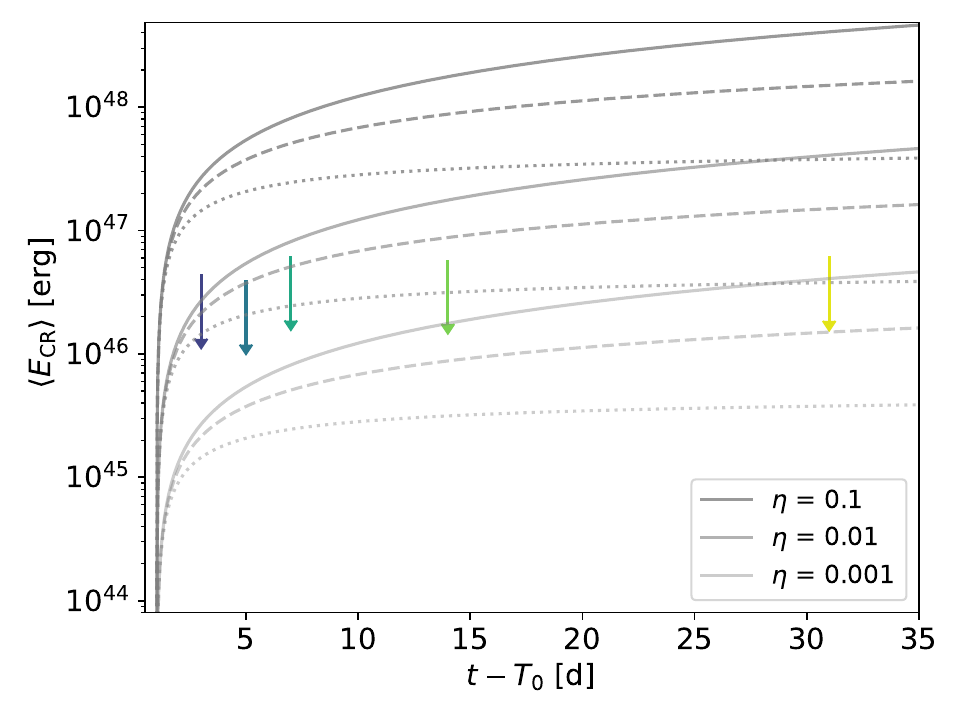}
   \caption{Limits for a steady-wind profile. (\textit{Left}): Likelihood profiles for different exposures considering the average density derived from a $\rho_w$ steady-wind profile with $m=1$, rescaling Fig. \ref{Fig:Likelihood_profile}, right panel. (\textit{Right}): Limits on the average cumulated total CR energy $\langle E_{\rm CR} \rangle$ for different exposure times $\Delta T$. $\langle E_{\rm CR} \rangle$ is computed for different efficiencies $\eta$ and velocity profiles with $m=1$ (solid line), $m=0.83$ (dashed line, as for SN 1993J), and $m=0.5$ (dotted line) using $\dot{M}_{\rm RSG} = 10^{-2}$ M$_{\odot}$/yr and $u_w = 100$ km/s. For visualisation purposes, \textit{Fermi}-LAT limits are only plotted for $m=1$. Colours represent the same exposures as in Fig. \ref{Fig:Likelihood_profile}.}
     
         \label{Fig:CR_limit}
  \end{figure*}

Nevertheless, a realistic density profile is not flat, but will likely follow the wind density of the progenitor \citep[see e.g.][]{Tatischeff09}. This can be firstly approximated as

\begin{equation}
                \rho_w(r) = \frac{\dot{M}_{\rm RSG}}{4 \pi r ^2u_w} \;. \label{eq:wind_profile}
\end{equation}

\noindent The SN shock will travel through this medium, and its radius $R_s(t)$ is typically parametrised during the free-expansion phase using the expansion parameter $m$ as

\begin{equation}
                R_s(t)= V_{s, 0} \left[ \frac{t}{1 \rm{day}}\right]^{m},
\end{equation}

\noindent where $V_{s, 0}$ is the initial shock velocity. The resulting one-dimensional density profiles for different $m$ values are shown in Fig. \ref{Fig:rho_profile}. Using those, we can rescale our likelihood limits, providing more realistic results, by using the average gas density for each exposure. This will lead to more stringent $E_{CR}$ constraints at earlier times due to the significantly larger densities closer to the stellar surface of the progenitor (see Fig. \ref{Fig:CR_limit}, left panel). However, we note that such a steady-wind density profile diverges at $T_0$, and therefore we only employ our limits integrating LAT data after $t_0=T_0 + 1$ d.

The energy-conversion efficiency can be finally parametrised through the parameter $\eta$ in the cumulative energy $E_{\rm CR}$ (i.e. as a fraction of the shock's kinetic energy), which is estimated as

\begin{equation}
                E_{\rm CR} = \eta \int ^{t_f} _{t_i} \frac{1}{2} \rho _w V_s ^3 4 \pi R_s^2 dt \; ,
\end{equation}

\noindent where $V_s = \frac{dR_s}{dt} = V_{s, 0} \left[ \frac{t}{1 \rm{day}}\right]^{m-1}$. We can compare our observations with the average cumulative $E_{\rm CR}$ over the different observing windows discussed in Section \ref{sec:data}, leading to an efficiency constraint at $\eta \lesssim 1\%$ (Fig.~\ref{Fig:CR_limit}, right panel). 

This efficiency limit is, in itself, a strong statement. Nevertheless, it depends on (1) our  assumed isotropic one-dimensional density profile, (2) a lack of $\gamma$-ray absorption, and (3) the presence of an average proton population accurately representing the true time evolution of the CRs for certain exposures. The reliability of these aspects is further discussed in Section \ref{sec:bias}.

\subsection{Galactic novae and their acceleration efficiency}

In order to explore the implications of our limits, we can compare the acceleration efficiency of SN with that of Galactic novae. The majority of novae detected at $\gamma$-rays originate in binary systems without an evolved stellar companion \citep[i.e. classical novae; see e.g.][]{Ackermann14, Chomiuk21}, and thus the surrounding conditions more  closely resemble those of Type Ia SNe. In those classical novae, shocks are found to be internal, displaying a strong correlation between optical and $\gamma$-ray flares \citep[][]{Aydi20}. The same behaviour arises in radiative shocks, where radiative cooling dominates the evolution of the ejecta. In such cases, the ratio between the optical and $\gamma$-ray luminosities should follow the relation \citep[see e.g.][]{Metzger15}

\begin{equation}
                \frac{L_{\gamma}}{L_{\rm opt}} = \frac{\eta_{p}}{\eta_{\gamma}} \;
,\end{equation}

\noindent where $\eta_{p}$ and $\eta_{\gamma}$ are the particle acceleration and $\gamma$-ray production efficiencies, respectively. If the shock is not fully radiative, the correlation instead provides a lower limit on the acceleration efficiency.

Despite the more energetic and rapidly expanding shock in SN 2023ixf, the environment in Type II SNe should be relatively similar to that in symbiotic novae, where an adiabatic shock travels through the wind of an RSG companion \citep[V407~Cyg and RS~Oph are the only unambiguously $\gamma$-ray-detected symbiotic binaries;][]{Abdo10,Cheung22}. Naturally, this analogy has its limitations, because the novae ejecta can be bipolar and the surrounding material inhomogeneous \citep{Munari22, Diesing23}. For our discussion, the case of RS Oph is particularly interesting as a relatively high efficiency ($\sim 10 \%$) was required to explain the VHE emission in the favoured hadronic scenarios \citep[see e.g.][]{Acciari22, HESS22}. We also note, that in this system, both external and internal shocks contribute to the emission \citep{Cheung22}, with a strong correlation between the optical and $\gamma$-ray luminosities of $L_{\gamma}/L_{\rm opt} \sim 2.5 \times 10^{-3}$ starting the day after the explosion. This ratio lies within the $10^{-4}$--$10^{-2}$ range of classical novae \citep[][]{Chomiuk21}. 

\begin{figure}
  \centering
   \includegraphics[width=\columnwidth]{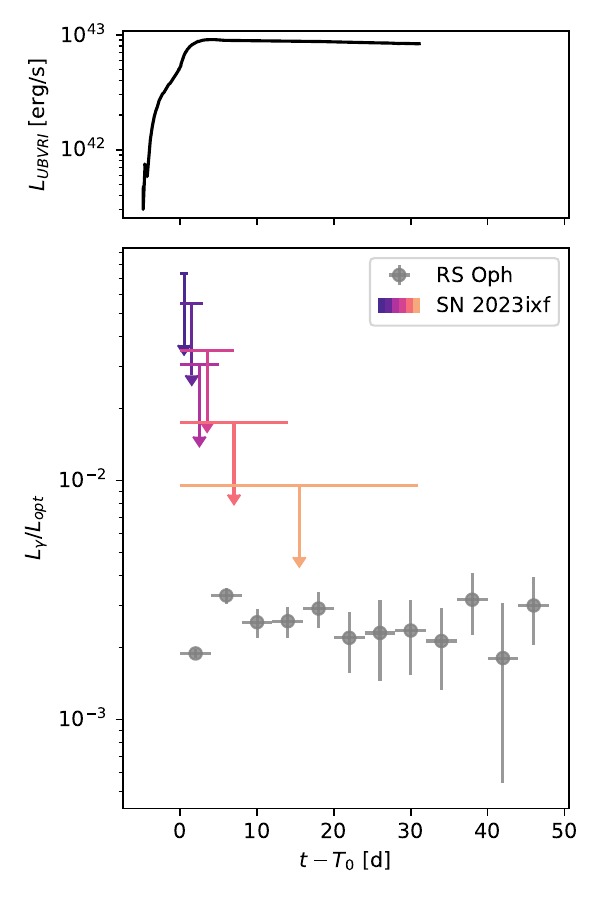}
   \caption{Luminosity of SN 2023ixf in the optical and GeV bands. (\textit{Top}): Example of a pseudobolometric luminosity ($L_{UBVRI}$) light-curve model of SN 2023ixf from Fig. 4 of \citet{Hiramatsu23}. We note that this is a conservative reference, as it is slightly lower than the bolometric luminosity derived by \citet{Bostroem23}, \citet{Teja23}, or \citet{Zimmerman23}, reaching $10^{43}$~erg/s. (\textit{Bottom}): Limits on the luminosity ratio between the optical and $\gamma$-ray bands. For comparison, the luminosity ratio from the symbiotic nova RS Oph is shown in grey \citep[][]{Cheung22}. Colours represent the same exposure times as in Fig. \ref{Fig:eflux}.}
     
         \label{Fig:lum_ratio}
  \end{figure}

This is in stark contrast to the low efficiency we derive in SN 2023ixf. Furthermore, we find that the luminosity ratio is of the same order ($\lesssim 1\%$; Fig.~\ref{Fig:lum_ratio}) as our spectral constraints on the efficiency. This provides an additional limit in case of the presence of a radiative shock. Either way, the relative weight of the non-thermal processes and particle acceleration in SN 2023ixf over its thermal electromagnetic output does not considerably exceed that of regular novae.

\subsection{Testing the shock breakout in $\gamma$-rays}

In the previous discussion, we centre our attention at $t > 1$~d for a shock propagating through the surrounding medium. However, our limits during the first day also provide a constraint on CR acceleration prior or close to the shock breakout, possibly characterised by a flash of UV/X-ray photons \citep[e.g. see the case of SN 2008D;][]{Chevalier08, Mazzali2008, Soderberg08}. This will occur at an optical depth of $\tau \sim c/V_s$, when a radiation-dominated shock travelling through the progenitor reaches the outer layers of the stellar component \citep[see][for a review]{Waxman17}. For a progenitor with an optically thin wind, the radiation-dominated shock precedes the formation of the collision-less, matter-dominated shock after the breakout. However, for dense, optically thick winds, the matter-dominated shock may form earlier than the shock breakout as photons are seized downstream up to a larger radii. In such a scenario, particle acceleration can occur as the collision-less shock forms deep into the wind if the condition

\begin{equation}
V_s \lesssim 4.3 \times 10^{4} \left[\frac{R_{\rm RSG}}{410 \rm{\;R_{\odot}}} \right] \left[\frac{\dot{M}_{\rm RSG}}{ 10^{-2} \rm{\;M_{\odot}/yr}} \right]^{-1} \left[\frac{u_{\rm{w}}}{100 \rm{\;km/s}} \right] \; \rm{km/s} \; 
\end{equation}

\noindent is fulfilled \citep{Giacinti15} \citep[we assume the RSG radius derived by][]{Hosseinzadeh23}. Therefore, for $V_s = 10^4$~km/s, CR acceleration could occur prior to the shock breakout up to several TeV. Although $\gamma$-rays produced through $\pi^0$ decay could be partially reprocessed within the shock into electron--positron pairs, GeV photons will likely not be absorbed (see Section \ref{sec:absorption}).

\citet{Li23} suggest that, for SN 2023ixf, the shock breakout might have occurred within a few hours after $T_0$. Our null result in the search for photon clusters in Section \ref{sec:clusters} discards the presence of a putative short, bright  flash emission in the $\gamma$-ray domain during the early expansion of the shock. In a complementary manner, our limit at $L_{\gamma} (E> 100 \; \rm{MeV}) < 4.8 \times 10^{41}$~erg/s for the first day of exposure can also be used to constrain the number of CRs accelerated at the time of the shock breakout.

\begin{figure*}[t]
  \centering
   \includegraphics[width=\columnwidth]{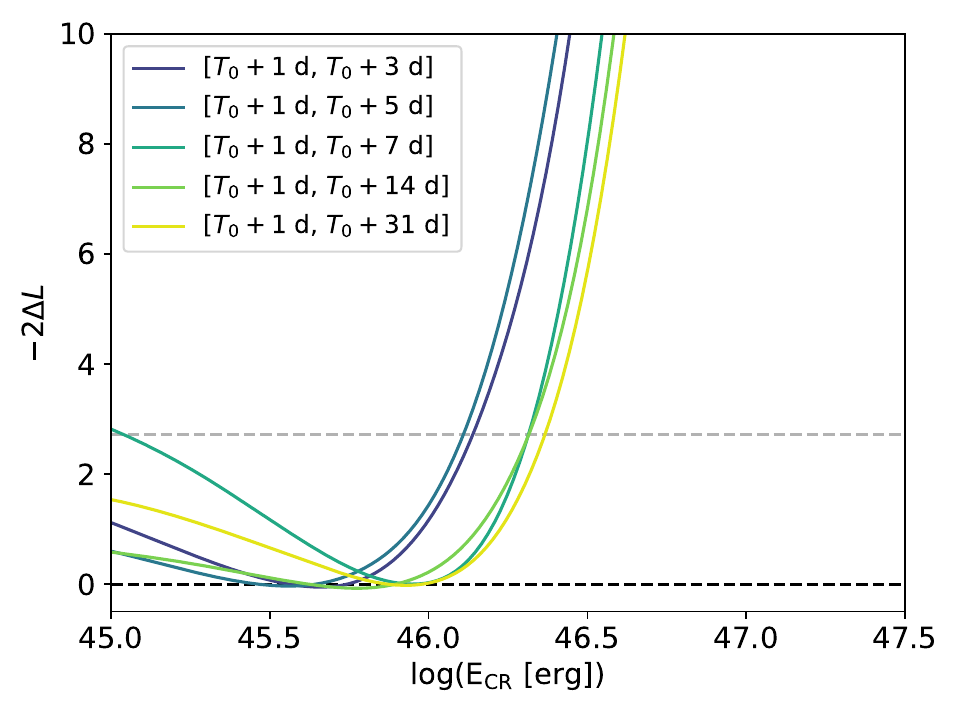}
   \includegraphics[width=\columnwidth]{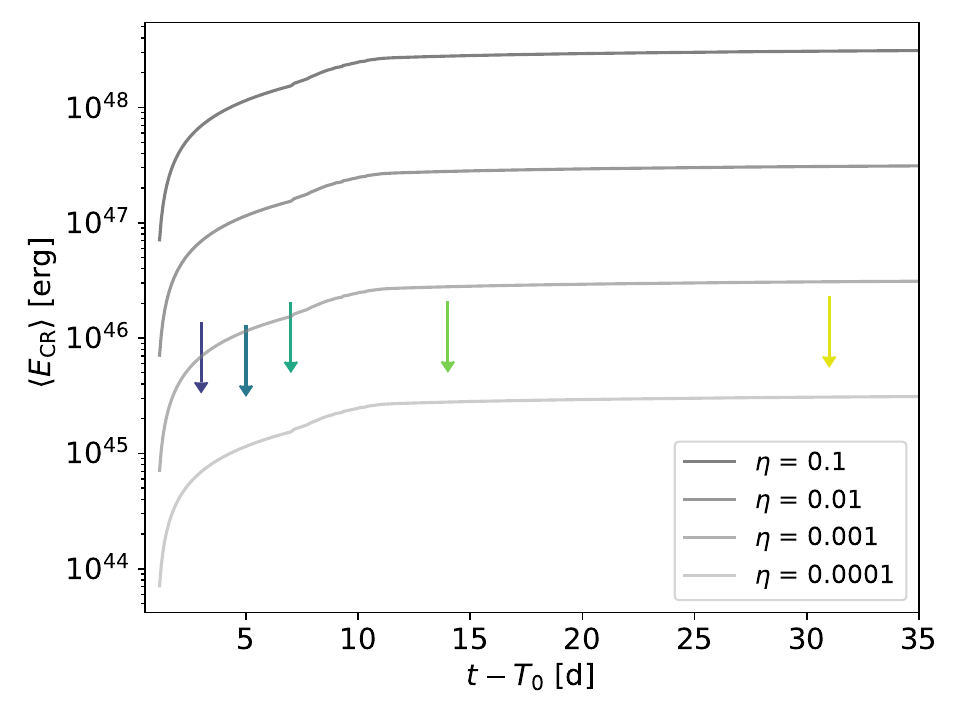}
   \caption{Limits using the best-fit density profile from \cite{Jacobson-Galan23} shown in Fig. \ref{Fig:rho_profile} (r1w6b model). (\textit{Left}): Likelihood profile obtained by rescaling those displayed in Fig. \ref{Fig:Likelihood_profile}, right panel. (\textit{Right}): Limits on the cumulated total CR energy derived as in Fig. \ref{Fig:CR_limit} but employing the aforementioned density profile.}
     
         \label{Fig:CR_limit_biases}
  \end{figure*}
  
\section{Possible biases} \label{sec:bias}

Although limits on the CR energy fraction at $10\%$ would be consistent with the standard Galactic CR origin paradigm, the new limits derived down to $1\%$ considerably quench the efficiency of SNe, and thus need to be considered with manifest scepticism. These limits were derived with reasonable but simplified assumptions, which may or may not apply. For example, the previously derived CR limit assumes an isotropic density profile with a constant wind before the explosion. Other assumptions affecting the underlying proton distribution, such as the presence or not of non-linear DSA, are likely to increase the photon flux at GeV in order to make SNe substantially contribute to the knee of the CR spectrum. For example, magnetic field amplification, which is required to retain CRs up to $E_p \sim 1$ PeV at early times, will soften the proton spectrum \citep[for a review see e.g.][]{Schure12, Blasi13}. Furthermore, it should be noted that our spectral predictions were made assuming quasi-steady-state equilibria at different times, which may not accurately represent the time evolution of the relativistic particle population during the shock expansion  \citep[see e.g.][]{Sarmah23}.

Among the possible biases, we further discuss the two relevant processes that, if wrongly interpreted, might underestimate the efficiency by orders of magnitude: uncertainties on the surrounding densities and $\gamma$-ray absorption processes. Additionally, we briefly discuss the timescales of the most relevant CR energy losses.

\subsection{The issue of the density profile and its homogeneity}

Generally, a density profile surrounding the progenitor can be described by a function of the form

\begin{equation}
                \rho_w = \rho_c \left( \frac{r}{r_c}\right) ^{-s} \; ,
\end{equation}

\noindent where a radially decreasing density has a characteristic value of $\rho_c$ at a distance $r_c$. Circumstellar medium (CSM) in the form of a steady wind will lead to $s = 2$ (see Eq. \ref{eq:wind_profile}), but larger values should be obtained for variable mass-loss rates. Optical observations consistently show that the progenitor was surrounded by a high-density CSM of $r <(0.3$ -- $1.0) \times 10^{15}$ cm \citep[][]{Smith23, Teja23, Bostroem23, Jacobson-Galan23, Zimmerman23}. This roughly corresponds to the distance that the progenitor's wind could reach in $\sim 1$~yr, while a $m=1$ shock travelling at $10^4$ km/s would need between one and two weeks to cross that high-density environment. At larger distances, the density is traced by the mass-loss rate of the stellar wind during the earlier pre-SN stages (likely closer to the typical value of $\dot{M}_{\rm RSG} = 10^{-6}$ M$_{\odot}$/yr). That is, there would be a transition with a steeper density profile $s>2$, which is not characterised by our assumptions in Fig. \ref{Fig:CR_limit}. However, conservatively, our early limits during this first week should be robust.

Importantly, these limits are still highly dependent on two observationally constrained quantities: $\dot{M}_{\rm RSG}$ and $u_w$. In Section \ref{sec:discussion}, we assumed $\dot{M}_{\rm RSG} = 10^{-2}$ M$_{\odot}$/yr and $u_w=100$ km/s. For ease of discussion, we define here the ratio $\omega = \dot{M}_{\rm RSG} /u_w$, and hence $\omega =6.3\times10^{16}$ g/cm. Our assumption for $\omega$ is consistent with the reported results from spectroscopic observations of ionisation lines and multi-band photometric light curves (see Appendix \ref{B:ratios} for a summary of multi-wavelength $\omega$ constraints). 

To explore the impact of those aspects, we reproduce the limits on Fig. \ref{Fig:CR_limit} but now testing the best-fit density model from \citet{Jacobson-Galan23}, which characterises the compact CSM (see Fig. \ref{Fig:rho_profile}). As shown in Fig. \ref{Fig:CR_limit_biases}, assuming the r1w6b model from \citet{Jacobson-Galan23} would imply an even stricter limit on the particle acceleration efficiency. Instead, we could consider a steady-wind profile with the lower limit on $\omega$ derived from SMA radio/mm observations (230 GHz, between $T_0 + 2$ d and $T_0 + 19$ d) by \citet{Berger23}. Employing their most conservative limit when considering free-free absorption \citep{Weiler86, Chevalier98} could relax the tension on the efficiency constraint, leading to a limit at $\eta \sim 16\%$ (see Fig.~\ref{Fig:Light_curve}). However, this would be in tension with the results from optical spectroscopic and photometric observations, as larger $\omega$ ratios are preferred.

Nevertheless, the exact value of the mass-loss rate is uncertain. A lower $\omega$ ratio has been reported from the modelling of the H$\alpha$ luminosity \citep{Zhang23} or in absorption features from X-ray observations  \citep{Grefenstette23, Chandra23, Panjkov23}. To explain the apparent tension with the ratios derived from different methods, it has been suggested that the CSM could be inhomogeneous \citep{Berger23}. This scenario could be explained by (1) an asymmetric distribution of the surrounding material, such as a dense torus caused by a pre-SN binary interaction, as proposed by \citet{Smith23} \citep[with asphericity in the ejecta also supported by the polarisation measurements from][]{Vasylyev23}, or (2) a pre-SN effervescent zone model with dense clumps embedded in a lighter RSG stellar wind (e.g. with $\rho_{\rm clump} \sim 3000 \rho_w$), as proposed by \citet{Soker23}. 
In both scenarios, our $\gamma$-ray constraints on the CR population could also be relaxed with the presence of an inhomogenoeus medium. That is, if the target gas ---with a density as derived from optical observations--- were to only occupy a volume filling factor $f_V\sim0.1$, an efficiency of $10\%$ in the CR acceleration would, a priori, still be compatible with the $\gamma$-ray constraints.

\subsection{The relevance of $\gamma$-ray absorption} \label{sec:absorption}

Although it appears that the uncertainties in the material distribution might explain our results, we can also consider the possibility that a significant fraction of the $\gamma$-ray flux is locally absorbed. Pair-production of electron--positron pairs can attenuate the high-energy photon flux at the source, while injecting a relativistic leptonic population into the shock. 

The $\gamma + \gamma \rightarrow e^- + e^+$ channel is likely to be irrelevant in our case, as the densest photon field produced by the SN will peak at UV and optical wavelengths; thus, the energy threshold for the interaction will be $\gtrsim 100$ GeV \citep[see e.g.][]{Dermer2009}. Contrarily, Bethe-Heitler (BH) pair-production (e.g. $p + \gamma \rightarrow p + e^- + e^+$) could have a larger impact for a hydrogen-poor, high-metallicity CSM \citep{Fang20}. A priori, both primary CRs and secondary $\gamma$-rays could be affected by this process. We note that relativistic protons would only interact with the SN optical photon field for $E_p>1$~PeV \citep[][for a shock temperature $T\sim10^4$ -- $10^5$~K]{Cristofari20}, and therefore our proton spectrum should remain unaltered, while secondary GeV $\gamma$-rays might still interact even with thermal plasma nuclei. However, the medium surrounding a SN Type II is hydrogen rich, and consequently BH should also have a negligible impact on the observed SED below 100 GeV \citep[$\tau_{\gamma} \lesssim 1$; threshold estimated at the time of the optical peak following][and references therein]{Fang20}.

 Nonetheless, we can attempt to estimate the consequences of a putative substantial pair-production, given that secondary relativistic pairs also radiate through the inverse Compton (IC), synchrotron, and non-thermal bremsstrahlung processes \citep[see e.g.][]{Blumenthal70,Baring99}. If there is a suppression of 90\% of the $\gamma$-ray flux, the leptonic emission from the pairs should not exceed the X-ray flux associated with thermal bremsstrahlung \citep[$\sim 2.5 \times 10^{-12}$ erg/cm$^2$/s between 0.3 and 79 keV;][]{Grefenstette23}. As a first approximation, we assume that all the energy from the $\gamma$-ray photons is converted into pairs. We note that under this assumption their energy content will exceed that of the secondary electrons and positrons from meson decays (e.g. $\pi^{\pm}$) in hadronic showers \citep[which would have a ratio of $N_{e^{\pm}}/N_{\gamma}=0.53$ for energies larger than the rest mass of the $\pi^0$;][]{Kelner06}. Under these assumptions (and also considering a uniform magnetic field with $B<10$ G), our estimated secondary X-ray fluxes from IC, synchrotron, and bremsstrahlung processes lie well below the measured thermal X-ray emission at early times. Therefore, although our theoretical estimates rule out absorption as an important actor in SN 2023ixf, the presence of secondary emission remains formally unconstrained by current X-ray and $\gamma$-ray observations.

\subsection{The impact of proton energy losses and CR escape} \label{sec:losses}

Here we discuss if our assumption of neglecting the CR evolution ---in order to provide quasi-model independent limits--- was valid. To this end, we computed: (1) the acceleration time $t_{\rm acc}$, (2) the escaping time $t_{\rm esc}$, (3) the timescale of the proton-proton interactions $t_{\rm pp}$, and (4) the time when adiabatic losses dominate the CR energy gain $t_{\rm ad, th}$. In our derivation, we adopt the assumptions considered in \citet{Vink2020} for a free expanding shock ($m=1$). For the upstream reference magnetic field we considered a value of $B=10$~G, which is rather conservative at early times, as its strength will likely decay in proportion to $t^{-1}$ \citep[see e.g.][]{Tatischeff09}.

First of all, we estimated the time required for particles to accelerate up to $E_{\rm p} \sim 1$ PeV via DSA. Assuming that the effective mean free path of the protons is effectively as small as its gyroradius (i.e. Bohm diffusion) and that the compression ratio at the shock is $\chi = 4$ with a magnetic field perpendicular to its normal, 

\begin{equation}
t_{\rm acc} \sim \frac{0.9}{\delta} \left[ \frac{E_{\rm p}}{1\; \rm{PeV}}\right] \left[ \frac{V_s}{10^4\; \rm{km/s}}\right]^{-2} \left[ \frac{B}{10\; \rm{G}}\right]^{-1} \rm{d}
\label{eq:acc}
,\end{equation}

\noindent where $\delta$ parametrises the energy dependence of the CR diffusion (realistically, $\delta \lesssim 1$). As previously mentioned, magnetic field amplification is necessary in order to accelerate CRs up to $E_{\rm p} \sim 1$~PeV at early times \citep{Schure12}. This will also impact the leakage of CRs escaping the shock, for which Bohm diffusion will cause particles with 

\begin{equation}
E_{\rm p} \gtrsim 2.6 \left[ \frac{B}{10 \;\rm{G}}\right]\left[ \frac{V_s}{10^4 \;\rm{km/s}}\right]^2 \left[ \frac{t}{1 \;\rm{d}}\right] \rm{PeV}
\end{equation}

\noindent to escape the shock region. Here we assume that the diffusion length is a small fraction of the shock radius ($l_{\rm{diff}} \lesssim 0.1 R_s$). Therefore, after one day, the SN shock should be able to retain CRs up to the knee of the CR spectrum for such high magnetic fields.

 \begin{figure*}[h]
  \centering
   \includegraphics[width=\columnwidth]{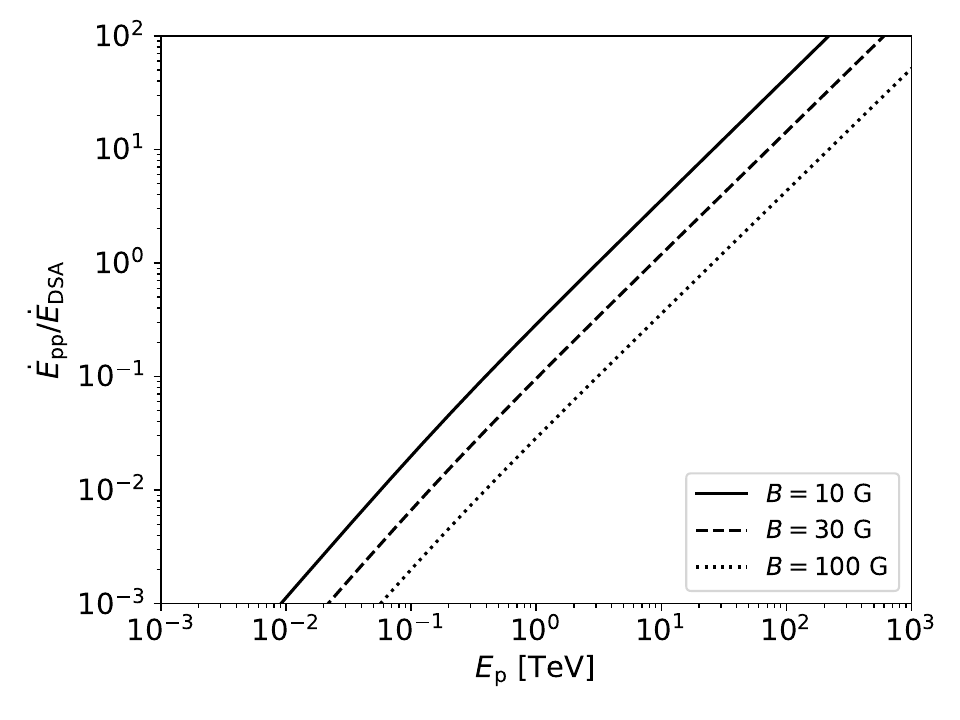}
   \includegraphics[width=\columnwidth]{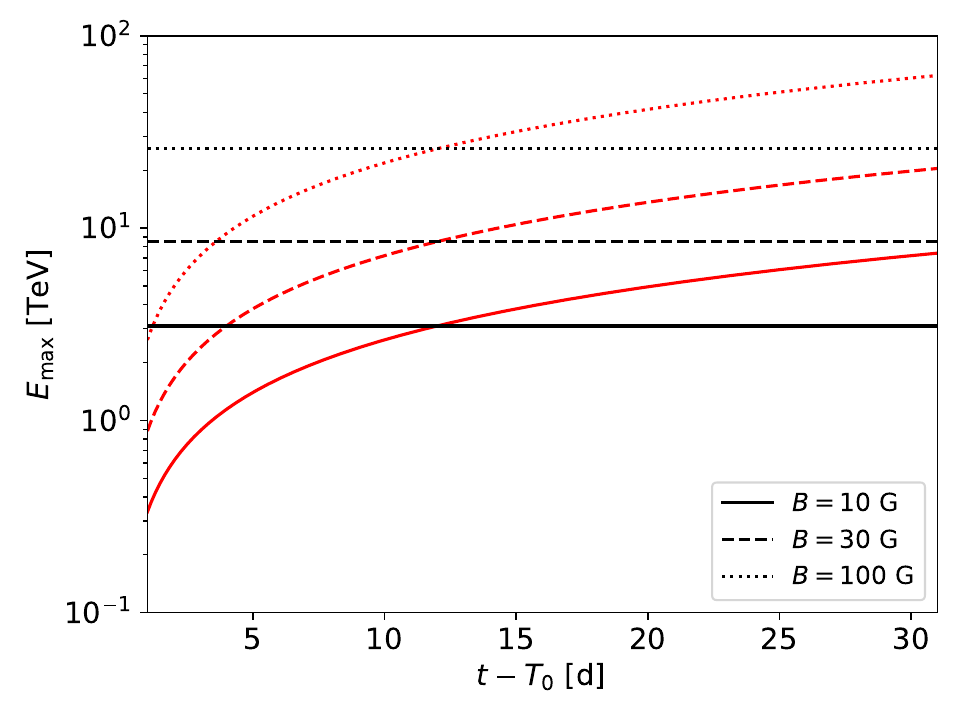}
   \caption{Impact of proton--proton losses on the maximum energy. (\textit{Left}): Ratio $\dot{E}_{\rm{pp}}/\dot{E}_{\rm{DSA}}$ as a function of the proton energy for $\rho _0$. (\textit{Right}): Maximum CR energy as limited by proton--proton interactions for a flat density profile with density ($\rho _0$, black) or a steady-wind profile (Eq. \ref{eq:wind_profile}; red), for different magnetic fields. $E_{\rm{max}}$ is derived by imposing $\dot{E}_{\rm{DSA}} = \dot{E}_{\rm{pp}}$ at the Bohm limit, and employing the parametrisation from \cite{Krakau15} for the proton--proton losses.}
         \label{Fig:emax_losses}
  \end{figure*}

However, for CR acceleration to be effective, we need the acceleration to occur on similar or shorter timescales than proton energy losses; that is, $t_{\rm{acc}} \lesssim t_{\rm pp}$
and $t_{\rm{acc}} \lesssim t_{\rm ad, th}$. For proton--proton interactions, we simply have 

\begin{equation}
t_{\rm pp} = \frac{\langle m \rangle}{\sigma_{pp} \rho_w(t) c} \; ,
\end{equation}

\noindent where $\langle m \rangle$ is the average mass for the particles in the CSM. For a hydrogen-rich environment with $\rho_w = \rho_0$, this implies $t_{\rm pp}\sim0.4$~d, which is similar to $t_{\rm{acc}}$. For this energy loss, only a small fraction of the CR energy is actually transferred to $\gamma$-rays ($\sim 5\%$), and it is therefore also similar to the DSA energy gain for a single scattering cycle ($\Delta E_p /E_p = \frac{4}{3}\frac{V}{c}\sim 4.4\%$).

Furthermore, we note that despite $\sigma_{pp}$ being quasi-energy independent at low energies, it does increase by a factor 2 at $1$~PeV \citep[see e.g.][]{Kafexhiu14}. Consequently, depending on the density, this energy loss could quench acceleration up to 1 PeV, but should not substantially affect protons with energies of less than a few TeV (see Fig.~\ref{Fig:emax_losses}), which still radiate in the GeV band. This in turn will only mildly impact the overall $E_{\rm CR}$ as, for example, an induced cut-off at 1 or 10~TeV reduces the total proton energy content by a factor of $\sim2$ (see Fig. \ref{Fig:SED_example}). Such a cut-off would be physically determined by the balance between the DSA energy gain ($\dot{E}_{\rm{DSA}}$) and the interaction losses ($\dot{E}_{\rm{pp}}$), while the overall normalisation of the proton population is instead set by the continuous CR injection at the shock. This, in turn, depends on the CSM geometry, which is the main source of uncertainty.
 
Finally, the free expansion of the shock will also lead to an adiabatic cooling of the CR population. Those energy losses will dominate over the DSA energy gain for any $t$ lower than

\begin{equation}
t_{\rm ad, th} \sim \frac{0.9}{\delta ^2} \left[ \frac{E_{\rm p}}{1\; \rm{PeV}}\right] \left[ \frac{V_s}{10^4\; \rm{km/s}}\right]^{-2} \left[ \frac{B}{10\; \rm{G}}\right]^{-1} \rm{d} \; ,
\end{equation}

\noindent where we assume an adiabatic index of $\gamma=4/3$ for the relativistic CRs. Consequently, adiabatic losses do not dominate the time evolution of the protons either after one day of the SN explosion. 

In view of those estimates and considering that we only integrate after $t=1$ d and neglect any possible prior larger emissivity, we do not expect a temporal-dependent description of the CR population to impact the derivation of our limits by an order of magnitude or more ---provided the underlying ideal shock conditions for effective particle acceleration hold.

\section{Summary} 
\label{sec:summary}

In the present study, we searched for $\gamma$-ray emission above 100~MeV from one of the closest CCSNe discovered since the \textit{Fermi} mission began, the Type II SN 2023ixf. We used standard likelihood-based analysis to look for $\gamma$-ray emission on different timescales starting from $T_0$ up to a month after the event. In a complementary manner, we employed a photon-counting algorithm to investigate photon clustering from the source position.

We do not detect a significant $\gamma$-ray signal from SN 2023ixf. Assuming (1) a simple, isotropic density profile for the CSM from a progenitor's mass-loss rate consistent with optical observations and (2) standard proton DSA up to 1 PeV, our observations imply a CR acceleration efficiency of $1\%$ or less. This result is in stark tension with the standard SN paradigm for the origin of Galactic CRs, which necessitates a 10\% efficiency for such a conversion process. As a first approximation, this tension can seemingly be alleviated assuming an inhomogeneous environment surrounding the progenitor. Therefore, a more sophisticated model  is required for both the shock and the CSM ---one that is consistent with all multi-wavelength observations. 

In essence, $\gamma$-ray observations are, for the first time, offering the opportunity to effectually constrain CR acceleration during the very early expansion of the shock produced by a CCSN event. To this end, we produced a comprehensive set of limits at high energies for the modelling of SN 2023ixf. Present and future \textit{Fermi}-LAT observations therefore provide a unique opportunity to establish whether or not SNe are indeed able to accelerate the bulk of CRs at early times up to the required energies.\\

\begin{acknowledgements}
The Fermi LAT Collaboration acknowledges generous ongoing support from a number of agencies and institutes that have supported both the development and the operation of the LAT as well as scientific data analysis. These include the National Aeronautics and Space Administration and the Department of Energy in the United States, the Commissariat \`{a} l'Energie Atomique and the Centre National de la Recherche Scientifique / Institut National de Physique Nucl\'{e}aire et de Physique des Particules in France, the Agenzia Spaziale Italiana and the Istituto Nazionale di Fisica Nucleare in Italy, the Ministry of Education, Culture, Sports, Science and Technology (MEXT), High Energy Accelerator Research Organization (KEK) and Japan Aerospace Exploration Agency (JAXA) in Japan, and the K. A. Wallenberg Foundation, the Swedish Research Council and the Swedish National Space Board in Sweden. Additional support for science analysis during the operations phase from the following agencies is also gratefully acknowledged: the Istituto Nazionale di Astrofisica in Italy and the Centre National d'Etudes Spatiales in France. This work performed in part under DOE Contract DE-AC02-76SF00515. G.P. acknowledges support by ICSC – Centro Nazionale di Ricerca in High Performance Computing, Big Data and Quantum Computing, funded by European Union – NextGenerationEU. Research at the Naval Research Laboratory is supported by NASA DPR S-15633-Y. \\

This work made use of Astropy:\footnote{\url{http://www.astropy.org}} a community-developed core Python package and an ecosystem of tools and resources for astronomy \citep{astropy:2013, astropy:2018, astropy:2022}, as well as Numpy\footnote{\url{https://numpy.org/}} \citep{numpy} and Matplotlib\footnote{\url{https://matplotlib.org/stable/}} \citep{Hunter07}. We are also thankful to the anonymous referee and our colleagues from the Fermi LAT Collaboration Philippe Bruel, Melissa Pesce-Rollins, Anita Reimer, Olaf Reimer, and David J. Thompson for their comments and suggestions on this work.

\end{acknowledgements}


\bibliographystyle{aa}
\bibliography{LAT_CR_SN2023ixf.bib}

\onecolumn
\begin{appendix}
\section{Summary of \textit{Fermi}-LAT upper limits} \label{A:limits}

Tables summarising all limits for different exposure times and energies. Flux and spectral upper limit results will be publicly accessible at \url{https://github.com/marti-devesa}.

\begin{table*}[h!]
\caption{Upper limits at 2$\sigma$ confidence level on SN 2023ixf for different exposure times $\Delta T$ starting at $T_0$. Integrated photon ($F_{\rm ph}$; in photons cm$^{-2}$ s$^{-1}$) and energy ($F_{\rm e}$; in erg cm$^{-2}$ s$^{-1}$) flux limits assume a spectral photon index $\Gamma=2$ and are derived as explained in Section \ref{sec:data}.}      
\label{table:ulims}      % is used to refer this table in the text
\centering                          % used for centering table
\begin{tabular}{c c c c c c c c}        % centered columns (4 columns)
\hline\hline      % inserts double horizontal lines
Time & TS & $F_{\rm ph}(>100 \;\rm{MeV})$ &$F_{\rm e}(>100 \;\rm{MeV})$ & $F_{\rm ph}(>1 \;\rm{GeV})$ &$F_{\rm e}(>1 \;\rm{GeV})$ & $F_{\rm ph}(>10 \;\rm{GeV})$ &$F_{\rm e}(>10 \;\rm{GeV})$ \\
%(photon cm$^{-2}$ s$^{-1}$) & & Photon flux (photon cm$^{-2}$ s$^{-1}$) & \\  
\hline                  % inserts single horizontal line
$T_0$ to 1 d & 0.0 & $6.3\times 10^{-8}$ & $8.6\times 10^{-11}$& $6.3 \times 10^{-9}$ & $5.8\times 10^{-11}$& $6.1 \times 10^{-10}$ & $3.5\times 10^{-11}$\\
$T_0$ to 3 d & 1.9 & $5.4\times 10^{-8}$ & $7.3\times 10^{-11}$& $5.3 \times 10^{-9}$ & $4.9\times 10^{-11}$ & $5.2 \times 10^{-10}$ & $3.0\times 10^{-11}$\\
$T_0$ to 5 d  & 0.9 & $3.1\times 10^{-8}$ & $4.2\times 10^{-11}$& $3.1 \times 10^{-9}$ & $2.8\times 10^{-11}$ & $3.0 \times 10^{-10}$ & $1.7\times 10^{-11}$ \\
$T_0$ to 7 d & 3.5 & $3.6\times 10^{-8}$ & $5.0\times 10^{-11}$& $3.6 \times 10^{-9}$ & $3.4\times 10^{-11}$ & $3.5 \times 10^{-10}$ & $2.0\times 10^{-11}$\\
$T_0$ to 14 d & 1.1 & $1.9\times 10^{-8}$ & $2.6\times 10^{-11}$& $1.9 \times 10^{-9}$ & $1.8\times 10^{-11}$& $1.8 \times 10^{-10}$ & $1.1 \times 10^{-11}$\\
$T_0$ to 31 d & 2.0  & $1.1\times 10^{-8}$ & $1.5\times 10^{-11}$  & $1.1 \times 10^{-9}$ & $1.0\times 10^{-11}$ & $1.1 \times 10^{-10}$ & $6.0\times 10^{-12}$ \\

\hline  %inserts single line
\end{tabular}
\label{values}
\end{table*}

\begin{table*}[h!]
\caption{Upper limits at 2$\sigma$ confidence level on SN 2023ixf for different exposure times $\Delta T$ starting at $T_1 = T_0+1$ d. Integrated photon ($F_{\rm ph}$; in photons cm$^{-2}$ s$^{-1}$) and energy ($F_{\rm e}$; in erg cm$^{-2}$ s$^{-1}$) flux limits assume a spectral photon index $\Gamma=2$ and are derived as explained in Section \ref{sec:data}.}      
\label{table:ulims}      % is used to refer this table in the text
\centering                          % used for centering table
\begin{tabular}{c c c c c c c c}        % centered columns (4 columns)
\hline\hline      % inserts double horizontal lines
Time & TS & $F_{\rm ph}(>100 \;\rm{MeV})$ &$F_{\rm e}(>100 \;\rm{MeV})$ & $F_{\rm ph}(>1 \;\rm{GeV})$ &$F_{\rm e}(>1 \;\rm{GeV})$ & $F_{\rm ph}(>10 \;\rm{GeV})$ &$F_{\rm e}(>10 \;\rm{GeV})$ \\
%(photon cm$^{-2}$ s$^{-1}$) & & Photon flux (photon cm$^{-2}$ s$^{-1}$) & \\  
\hline                  % inserts single horizontal line
$T_1$ to 3 d & 2.3 & $7.2\times 10^{-8}$ & $9.8\times 10^{-11}$ & $7.1 \times 10^{-9}$ & $6.6\times 10^{-11}$ & $6.9 \times 10^{-10}$ & $4.0\times 10^{-11}$ \\
$T_1$ to 5 d  & 1.2 & $3.6\times 10^{-8}$ & $4.9\times 10^{-11}$& $3.6 \times 10^{-9}$ & $3.3\times 10^{-11}$ & $3.5 \times 10^{-10}$ & $2.0\times 10^{-11}$\\
$T_1$ to 7 d & 4.0 & $4.0\times 10^{-8}$ & $5.5\times 10^{-11}$& $4.0 \times 10^{-9}$ & $3.7\times 10^{-11}$ & $3.9 \times 10^{-10}$ & $2.2\times 10^{-11}$\\
$T_1$ to 14 d & 1.2 & $2.0\times 10^{-8}$ & $2.7\times 10^{-11}$& $2.0 \times 10^{-9}$ & $1.8\times 10^{-11}$& $1.9 \times 10^{-10}$ &  $1.1\times 10^{-11}$\\
$T_1$ to 31 d & 1.9  & $1.1\times 10^{-8}$ & $1.4\times 10^{-11}$  & $1.1 \times 10^{-9}$ & $9.7\times 10^{-12}$ & $1.0 \times 10^{-10}$ & $5.8\times 10^{-12}$ \\

\hline  %inserts single line
\end{tabular}
\label{values}
\end{table*}

\begin{table*}[h!]
\caption{Upper limits on SN 2023ixf for different exposure times $\Delta T$.}      
\label{table:ulims}      % is used to refer this table in the text
\centering                          % used for centering table
\begin{tabular}{c c c c c c c c}        % centered columns (4 columns)
\hline\hline      % inserts double horizontal lines
Time & $L_{\gamma}$ (erg/s) &  $E_{\rm CR} \left[\frac{\rho}{\rho_0}\right]^{-1} $ (erg)\\
%(photon cm$^{-2}$ s$^{-1}$) & & Photon flux (photon cm$^{-2}$ s$^{-1}$) & \\  
\hline                  % inserts single horizontal line
$T_1$ to 3 d  & $5.5 \times10^{41}$ & $1.2 \times10^{47}$ \\
$T_1$ to 5 d  & $2.8 \times10^{41}$ & $7.6 \times10^{46}$ \\
$T_1$ to 7 d  & $3.1 \times10^{41}$ & $9.3 \times10^{46}$ \\
$T_1$ to 14 d  & $1.5 \times10^{41}$ & $4.5 \times10^{46}$ \\
$T_1$ to 31 d  & $7.9 \times10^{40}$ & $2.3 \times10^{46}$ \\

\hline  %inserts single line
\end{tabular}\\
\label{values}
\tablefoot{Limits on the CR energy are derived as explained in Section \ref{sec:discussion} and normalised to $\rho_0 =5.6\times 10^{-14}$ g/cm$^3$. Luminosities were obtained for a distance of $D=6.85$ Mpc \citep[][]{Riess22}. The impact of redshift and extragalactic background light absorption is negligible and has not been considered.}
\end{table*}

\begin{figure*}[h!]
  \centering
   \includegraphics[width=\columnwidth]{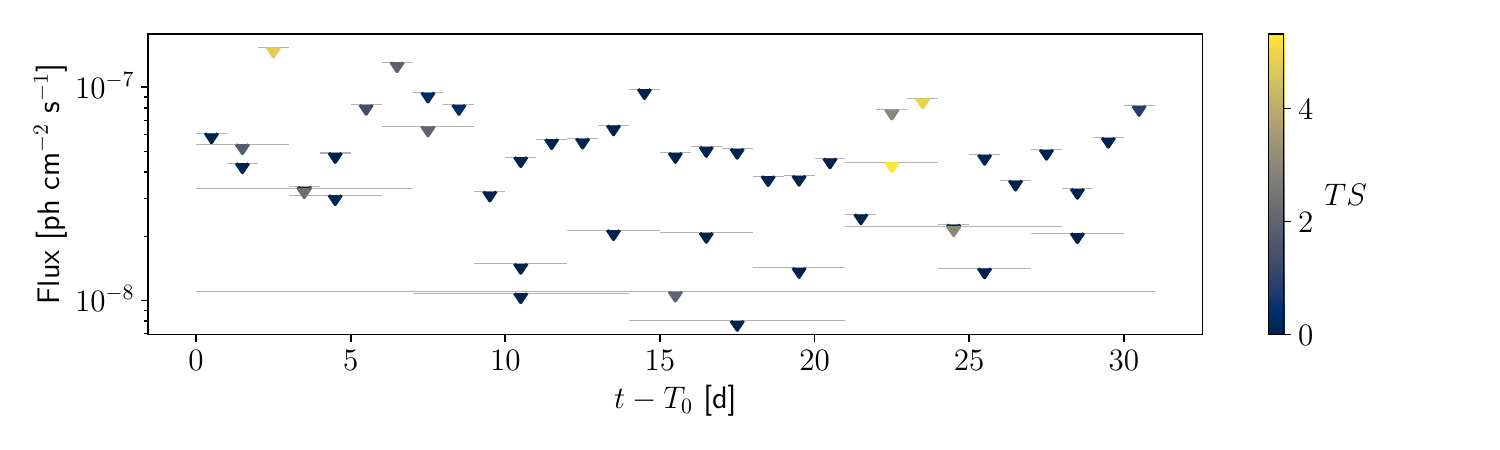}
   \caption{\textit{Fermi}-LAT upper limits for different time bins (1, 3, 7, and 31 days) during the whole month of observations. Flux limits assume a spectral photon index $\Gamma=2$ and are derived as explained in Section \ref{sec:results}.}
         \label{app:Light_curve}
  \end{figure*}

\clearpage

\section{Previous multi-wavelength constraints on the RSG pre-outburst stellar wind} \label{B:ratios}

\begin{table*}[h!]
\caption{Summary of multi-wavelength constraints on the surrounding material around the progenitor from previous works.}             %title of Table
\label{table:etacar}      % is used to refer this table in the text
\centering                          % used for centering table
\begin{tabular}{c c c c c}        % centered columns (4 columns)
\hline\hline      % inserts double horizontal lines
Method & $\dot{M}_{\rm RSG}$ (M$_{\odot}$/yr) & $u_w$ (km/s) & $\omega =\dot{M}_{\rm RSG}/u_w$ (g/cm) & Publication \\  
\hline                  % inserts single horizontal line
Radio/mm lower limit$^a$ & $>7\times 10^{-4}$ and $>3.5\times 10^{-3}$ & $115^b$ & $>3.8\times10^{15}$ & \citet{Berger23}\\
Spectroscopy &  $10^{-3}$ -- $10^{-2}$ & $150$ & $\sim 2.1\times10^{16}$ & \citet{Bostroem23} \\
X-ray absorption & $3\times 10^{-4}$ & $50$ & $3.7\times10^{15}$ & \citet{Grefenstette23}\\
Multi-band photometry & $10^{-2}$ -- $10^{-0}$ & $115^b$ & $\sim 5.5\times10^{17}$ & \citet{Hiramatsu23}\\
Multi-band photometry & $10^{-2}$& $50$ & $1.3\times10^{17}$ & \citet{Jacobson-Galan23} \\
Optical photometry& $10^{-3}$ & $10$ & $6.3\times10^{16}$ & \citet{Teja23}  \\
 H$_{\alpha}$ luminosity & $6\times 10^{-4}$ & $55$ & $6.9\times10^{15}$  & \citet{Zhang23}\\

Bolometric luminosity & $1.1\times 10^{-2}$ & $30$ & $2.3\times10^{17}$  & \citet{Zimmerman23}\\

\hline

-- & $10^{-2}$ & $100$ & $6.3\times10^{16}$ & Assumption in this work\\

\hline  %inserts single line
\end{tabular}
\label{values}
\tablefoot{For those works reporting a range of possible mass-loss rates, an intermediate value is provided for the ratio. For those works reporting a range of limits, the most conservative one is used as a reference for the ratio.\\
\tablefoottext{a}{Obtained from Fig. 3 in \citet{Berger23} for $T_e=10^4$ K and $T_e=10^5$ K, respectively, assuming $V_s=10^4$ km/s. Valid for a CSM with a wind profile truncated at $2\times10^{15}$ cm.}\tablefoottext{b}{Derived in \citet{Smith23}.}}
\end{table*}
 
\end{appendix}
\end{document}